\documentclass[preprint, superscriptaddress]{revtex4-2}
\makeatletter
\def\@fnsymbol#1{\ensuremath{\ifcase#1\or *\or \mathsection\or \dagger\or \ddagger\or
    \mathparagraph\or \|\or **\or \dagger\dagger
   \or \ddagger\ddagger \else\@ctrerr\fi}}
\makeatother
\usepackage{bbold}
\usepackage{amsmath,amsfonts,mathtools,bbm,bm}
\usepackage{graphicx}
\usepackage{braket}
\usepackage{cases}
\usepackage[normalem]{ulem}
\usepackage[colorlinks,linkcolor=blue,urlcolor=blue,citecolor=blue]{hyperref}
\usepackage[all]{hypcap}
\usepackage{orcidlink}
\usepackage{xcolor}
\usepackage{tikz}
\usepackage{cases}
\usetikzlibrary{arrows.meta,decorations.pathreplacing,calc,patterns}
\newcommand{\beq}{\begin{equation}}
\newcommand{\eeq}{\end{equation}}
\usepackage{cancel}
\begin{document}
\title{Planckian dissipation from classical hydrodynamics}
\author{Laura Foini~\orcidlink{0009-0007-4906-5533}}
\email{laura.foini@ipht.fr}
\affiliation{IPhT, CNRS, CEA, Universit\'e Paris Saclay, 91191 Gif-sur-Yvette, France}
\author{Jorge Kurchan~\orcidlink{0000-0001-5367-3688}}
\affiliation{Laboratoire de Physique de l'\'{E}cole Normale Sup\'erieure, ENS, Universit\'e PSL, CNRS,
Sorbonne Universit\'e, Universit\'e de Paris, F-75005 Paris, France}
\author{Silvia Pappalardi~\orcidlink{0000-0001-6931-8736}}
\affiliation{Institut f\"ur Theoretische Physik, Universit\"at zu K\"oln, Z\"ulpicher Stra{\ss}e 77, 50937 K\"oln, Germany}
\date{\today}
\begin{abstract}
Hydrodynamics describes the long-time, long-distance dynamics of interacting quantum systems by classical equations, yet at low temperature quantum fluctuations might eventually invalidate this classical picture. Here we ask what it takes for a quantum system to sustain a consistently classical hydrodynamic description.
Our starting point is the quantum fluctuation–dissipation theorem which, when read in the time domain, acts as a blurring of the fine details of correlation functions on a Planckian time-scale. We analyse how this blurring interacts with the diffusive large-deviation form of correlators assuming the existence of a hydrodynamic light cone velocity, and show that the interior of the cone splits into a classical region, where correlation and response satisfy the classical fluctuation–dissipation relation, and a quantum region, where they deviate sharply from it. We illustrate the mechanism on three phenomenological hydrodynamic equations — diffusion, telegraph and diffusive-telegraph. Preserving a finite classical region as the temperature is lowered forces the current relaxation rate to be bounded from above by the Planckian rate. This bound connects to bounds on diffusivity, equilibration time and shear viscosity. In this way, a Planckian bound on dissipation emerges not as a quantum constraint on microscopic dynamics, but as the price a system pays to remain describable by classical hydrodynamics down to low temperatures.
\end{abstract}
\maketitle
\section{Introduction}
A central puzzle in quantum many-body physics is why so many strongly correlated materials~\cite{hartnoll2022colloquium,chowdhury2022sachdev} -- from cuprates to heavy fermions to cold-atom gases -- relax on a timescale set only by temperature and Planck's constant,
\beq
\tau_{\mathrm{Pl}} \;=\; \frac{\hbar}{k_B T}\ .
\eeq
Whenever transport, thermalisation or chaos takes place on a time of order $\tau_{\mathrm{Pl}}$, one speaks of a \textit{Planckian bound}~\cite{zaanen2004temperature,hartnoll2022colloquium}. The appearance of $\hbar$ and $k_B T$, without any material-specific constant, hints at a universal minimal timescale.

Several incarnations of this idea have been discussed in the literature. On the transport side, the Kovtun–Son–Starinets bound $\eta/s\ge \hbar/(4\pi k_B)$~\cite{kovtun2005viscosity} sets a universal lower limit on shear viscosity over entropy density. On the chaos side, Maldacena–Shenker–Stanford proved that the quantum Lyapunov exponent obeys $\lambda_L \le 2\pi k_B T/\hbar$~\cite{maldacena2016bound}. 
More recently, bounds on the local equilibration time have been proposed~\cite{hartman2017upper,delacretaz2025bound, qi2026planckian}, and a Planckian bound on the dynamical entropy has also been conjectured~\cite{cao2025planckian}. Such planckian bounds have also been derived from quantum uncertainty relations~\cite{nussinov2022exact,nussinov2025planckian} and put in relation with the crossover between laminar and turbulent flow in superfluids~\cite{volovik2022quantum}.

Despite this abundance of examples, what actually \textit{enforces} these bounds at the microscopic level has remained elusive. A large body of work has sought concrete mechanisms producing Planckian, $T$-linear scattering rates: universal diffusion bounds for incoherent metals~\cite{hartnoll2015theory}, solvable Sachdev--Ye--Kitaev-type models~\cite{chowdhury2022sachdev} of strange metals at the breakdown of a heavy Fermi liquid~\cite{aldape2022solvable} or with spatially random interactions~\cite{patel2023universal,patel2025strange}, incoherent carriers scattered by local moments~\cite{ciuchi2023strange}, local polaron formation~\cite{lee2026apparent}, stability constraints on conventional metals~\cite{murthy2023stability} and information-theoretic optimality~\cite{chowdhury2026information}. Closest to the real-time perspective adopted here are non-perturbative real-space, real-time studies of Planckian resistivity~\cite{aydin2024quantum} and the ray-by-ray generalisation of the chaos bound inside the butterfly cone~\cite{mezei2020chaos}. 

A complementary route, which does not rely on any specific mechanism, uses only general properties of thermal correlation functions, as for the bound on chaos~\cite{maldacena2016bound} and the related bounds on two-point functions~\cite{delacretaz2025bound,qi2026planckian}. Along this line, it was shown in Refs.~\cite{tsuji2018bound,pappalardi2021quantum} that many known bounds, including the bound on chaos, follow as a direct consequence of the quantum fluctuation–dissipation theorem (FDT) once it is rewritten in the time domain. In that formulation, the quantum FDT relates the symmetrised correlator $C(x,t)$, the response $R(x,t)$ - or its integrated version $\Psi(x,t)$ - and the regulated correlator $F(x,t)$ - the object on which proofs of Planckian bounds usually rely~\cite{maldacena2016bound,qi2026planckian}, and which coincides with $C$ in the classical limit - through a \textit{blurring} of fine time-details on the intrinsic scale
\beq
\label{eq_defOmega}
\frac{1}{\Omega}\;=\;\frac{\hbar\beta}{\pi}=\frac{\tau_{\rm Pl}}{\pi} ,
\eeq
where $\Omega$ is the first Matsubara frequency. In the classical limit $\hbar\beta\to 0$ the blurring reduces to a Dirac delta, and one recovers the usual classical fluctuation-dissipation relation. At finite $\hbar\beta$, this blurring irons out structures shorter than the Planckian time, and it is precisely this operation that gives rise to bounds on how fast correlators can decay or grow.

The present work applies this perspective to \textit{hydrodynamics}. While we require long-time diffusive behaviour, we also assume the existence of a light cone with velocity $v$ that restricts the propagation of the conserved density within the causal region $x\leq vt$. The emergence of an effective light cone is generic in locally interacting many-body systems; in the quantum case it is formalised by the Lieb–Robinson bound~\cite{lieb1972finite}. Numerous analytical, numerical and experimental studies~\cite{calabrese2006time,lauchli2008spreading,manmana2009time,kim2013ballistic,cheneau2012light,langen2013local} have confirmed the robustness of such light-cone-like spreading. The tension between diffusion and finite-speed propagation also appears in relativistic hydrodynamics~\cite{baier2008relativistic,romatschke2010new} and has been used to bound diffusivity beyond that context~\cite{hartman2017upper} (see also \cite{chowdhury2025bootstrapping,
chowdhury2025lieb} on the constraints of the Lieb-Robinson velocity imposed on retarded Green's functions).

We stress that $v$ is meant here as a \textit{hydrodynamic} velocity -- the speed of reversible propagation of the conserved density -- rather than a microscopic kinematic speed. Following the logic of extended (Cattaneo--Israel--Stewart) hydrodynamics~\cite{israel1976nonstationary,baier2008relativistic,romatschke2010new}, its existence amounts to promoting the slow part of the current to a dynamical variable, which relaxes to Fick's law on a time $\tau_J$, with $v^2=D/\tau_J$. For this to be a hydrodynamic description, the current must be slow compared with all non-hydrodynamic modes, i.e.\ $\tau_J$ must exceed the time $\tau_{\rm hydro}$ beyond which all fluctuations are slaved to those of the conserved quantities and local equilibrium (a local temperature, a local density) is established:
\beq\label{eq:hydro_assumption}
\tau_{\rm hydro}\;\leq\;\tau_J\ .
\eeq
This is what distinguishes hydrodynamic transport from mere diffusion. 
This condition fails for instance in a disordered metal which is goverened by single-particle diffusion and where our arguments do not hold, as we better discuss in the conclusions.

The starting observation of our work is simple: hydrodynamics -- the coarse-grained dynamics of the few slow, conserved quantities -- is inherently a \textit{classical} description. Even if derived from a quantum microscopic theory, the resulting equations for density, momentum or energy do not carry any explicit factor of $\hbar$. The question we ask is: given a hydrodynamic equation, up to what temperatures can its diffusive solution be consistently viewed as classical correlators, i.e.\ as correlators for which the quantum FDT reduces to its classical form? 

As an illustration, we study the dynamics on rays $x=avt$ inside the light cone ($0<a<1$), using three representative phenomenological equations for a conserved density: the diffusion equation (instantaneous propagation), the telegraph equation (finite relaxation time, finite speed $v$, hard light cone) and the diffusive-telegraph equation (a softer version of the latter). In all of them the long-time behaviour of the correlator on a ray takes the form
\beq\label{ansatz_intro}
C(x=avt,t)\;\sim\; e^{-\lambda(a)\, t}\ ,
\eeq
with a rate $\lambda(a)$ that vanishes as $\lambda(a) = \lambda a^2$ on the time axis and grows as one approaches the light cone. Matching to the diffusive regime identifies $\lambda = v^2/(4D)$. The form~\eqref{ansatz_intro} is compatible with a quadratic small-$a$ structure of the large-deviation function, preserving an effectively diffusive (Gaussian) hydrodynamic core. The time-scale $\lambda^{-1}$ is set by the relaxation time of the current, $\lambda^{-1}=4\tau_J$; it is responsible for the emergence of a light-cone front, as in relativistic hydrodynamics~\cite{israel1976nonstationary}.

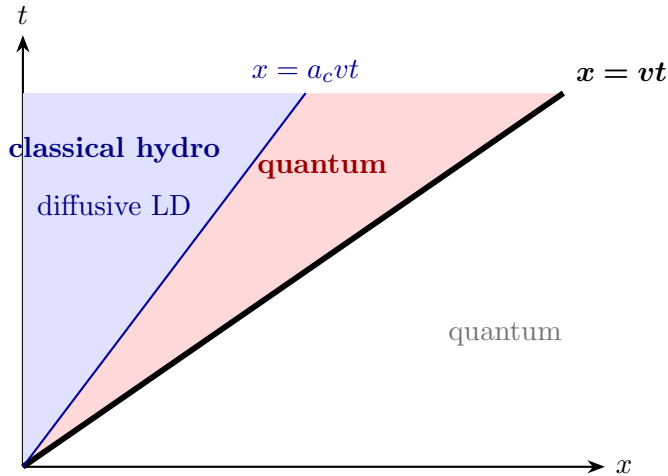
\begin{figure}[t]
\centering
\begin{tikzpicture}[scale=1.1, >=Stealth, thick]
  \draw[->] (0,0) -- (7.0,0) node[right] {$x$};
  \draw[->] (0,0) -- (0,5.2) node[above] {$t$};
  \fill[blue!12] (0,0) -- (3.4,4.5) -- (0,4.5) -- cycle;
  \fill[red!15]  (0,0) -- (3.4,4.5) -- (6.5,4.5) -- cycle;
  \draw[line width=2.2pt] (0,0) -- (6.5,4.5)
    node[above right] {$\boldsymbol{x=vt}$};
  \draw[thick,blue!60!black] (0,0) -- (3.4,4.5)
    node[above, blue!60!black] {$x=a_c vt$};
  \node at (1.1,3.5) [blue!50!black,align=center]
  {\textbf{classical hydro} \\ diffusive LD};
  \node at (3.6,3.3) [red!60!black,align=center] {\textbf{quantum} \\ };
  \node at (5.8,1.6) [black!55,align=center] {quantum};
\end{tikzpicture}
\caption{\label{fig:scheme}%
  Cartoon of the mechanism. Inside the hydrodynamic light cone $|x|<vt$, correlators decay exponentially along a ray $x=avt$ with a rate $\lambda(a)$ that vanishes on the time axis and grows toward the cone. Starting from a classical large-deviation ansatz, the quantum FDT blurs the fine time-details at the Planckian scale $\hbar\beta/\pi$. Inside the inner (blue) cone the blurring is irrelevant and different correlation functions that classically coincide at equilibrium remain so. Beyond this boundary the blurring dominates: the constraints imposed by the quantum FDT, the presence of the light cone, and diffusivity compete and the resulting dynamics is Planckian. Keeping the classical region finite as $\hbar\beta\to\infty$ forces $v^2/(4D)$ to be at most of order $\Omega$.}
\end{figure}

The resulting picture is summarised in Fig.~\ref{fig:scheme}. The Planckian blurring imposed by the quantum FDT is irrelevant whenever $C(t),F(t),\Psi(t)$ vary slowly: there $F\simeq C\simeq T\Psi$, and the form of hydrodynamics is unaffected by quantum fluctuations. Conversely, when~\eqref{ansatz_intro} enforces a rapid temporal variation, the blurring dominates: the other functions acquire a Planckian rate $\Omega$ in some spatio-temporal regime, very different from the diffusive growth originally postulated. In the models we consider, the boundary between the two regimes is a critical ray $x=a_c v t$, inside which diffusive hydrodynamics can be read classically. Three scenarios occur:
\begin{itemize}
    \item $a_c \to 1$ (when $\lambda^{-1} \gg \tau_{\rm Pl}$), the system is classical throughout the light cone;
    \item $a_c \to 0$ (when $\lambda^{-1} \ll \tau_{\rm Pl}$), the system is quantum for all rays;
    \item $a_c$ finite (when $\lambda^{-1} \sim \tau_{\rm Pl}$) a critical cone separates the two regimes.
\end{itemize}
Since $\lambda(a)\sim a^2 v^2/(4D)$ at small $a$, demanding that $a_c$ remain finite as $T\to 0$ imposes
\beq\label{eq:PlScaling_intro}
\frac{v^{2}}{4 D}\;\lesssim\;\Omega\;=\;\frac{\pi}{\hbar\beta}\ ,
\eeq
a Planckian bound, where $\lesssim$ is understood up to a temperature-independent prefactor: what is excluded is a divergence of $\lambda/\Omega$ as $T\to0$, as would occur for a temperature-independent $\lambda$. The main novelty of the present perspective is that this  bound is \textit{not} an extra quantum constraint imposed on top of hydrodynamics, but the minimal requirement for diffusive hydrodynamics to remain classical at low temperature in a finite region of space-time.

The paper is organised as follows. In the Results section we derive, in the simplest possible terms, the mechanism that produces the Planckian scaling~\eqref{eq:PlScaling_intro}, then specialise to three phenomenological hydrodynamic equations -- diffusion, telegraph and diffusive-telegraph -- and present numerical results illustrating the mechanism on the telegraph equation. The Discussion places our findings in the context of other Planckian bounds. The Methods section recalls the definitions of the relevant correlators, states the quantum FDT in the time domain, and explains its interpretation as a blurring. Appendices collect complementary FDT identities, a direct computation of the fluctuation-dissipation ratios on rays, and a self-contained summary of the three phenomenological hydrodynamic theories used in the main text.

\section{Results}

\subsection{The mechanism: when hydrodynamics stays classical}\label{Sec_general}

Let us see how the blurring imposed by the quantum FDT (see Methods) interacts with hydrodynamics and how it generates a Planckian scaling. The specific form of the hydrodynamic equation is unimportant at this level: we treat it in generality and only later, in Sec.~\ref{Sec_eqs}, specialise to concrete examples. We restrict the discussion to one spatial dimension; the argument generalises immediately.

\textbf{A decay rate that depends on where you look.}
Consider a conserved density $n(x,t)$ -- which could be a transverse velocity if we are studying shear -- and its symmetrised connected correlator $C(x,t)=\langle n(x,t)n(0,0)\rangle_{\rm sym}$. We assume two generic features. First, a \textit{finite propagation speed} $v$: the response cannot spread faster than some effective velocity, so the response function is essentially confined to the light cone $x<vt$ (and so is $C$, up to a Planckian `leakage' discussed below). As discussed in the Introduction, $v$ is here the hydrodynamic velocity associated with a current that relaxes on a time $\tau_J$, slow compared with local equilibration, Eq.~\eqref{eq:hydro_assumption}. Second, \textit{diffusive dynamics at late times}: inside the cone, at times much longer than any microscopic scale, the density diffuses with constant $D$.

A convenient parametrisation that captures both properties is the \textit{large-deviation ansatz}:
\beq\label{eq:LDP_ansatz}
 C(x, t) \simeq\; \frac 1{\sqrt t} e^{-\lambda\, t\,\Phi\bigl(\frac{x}{vt}\bigr)}\ ,
\eeq
with $\Phi(0)=0$, $\Phi(a)\sim a^{2}$ for $a\to 0$ (so that the diffusive Gaussian is recovered near the time axis) and $\Phi(a)$ growing toward the light cone $a=1$. The overall rate is fixed by matching to diffusion:
\beq\label{eq:lambda_vs_D}
\lambda\;=\;\frac{v^{2}}{4 D}\ .
\eeq
At fixed $x$, Eq.~\eqref{eq:LDP_ansatz} describes the growth of the diffusive peak after the light cone. At fixed ray $a=x/(vt)$, the correlator decays exponentially in time as
\beq\label{eq:Cray}
C(x=avt,\,t)\;\sim\; e^{-\lambda \Phi(a)\, t}\ .
\eeq
The physical intuition is simple: \textit{the rate at which correlations decay depends on how close one is to the light cone}. Near the time axis ($a\to 0$) the decay is slow and diffusive, $\lambda(a)\simeq \lambda a^{2}$; closer to the cone ($a\to 1$) it becomes faster, because fluctuations there are sensitive to the inertial, wave-like part of the dynamics. The ansatz~\eqref{eq:LDP_ansatz} is classical in nature: $\hbar$ does not appear explicitly. Under stochastic dynamics, the same large-deviation form emerges naturally in diffusive systems with a light cone, since conditioning on atypical displacements $x\simeq t$ probes exponentially rare fluctuations beyond the typical $x\simeq\sqrt{t}$ spreading.

\textbf{Blurring versus hydrodynamic decay.}
Given $C(x,t)$, we can ask how $F$ and $\Psi$ behave on the same ray. By Eqs.~\eqref{eq:blurring} in Methods, both are obtained from $C$ through convolution with a positive kernel of width $\Omega^{-1}=\hbar\beta/\pi$ (the kernel $g_\Omega$ for $F$ and $D_\Omega$ for $T\Psi$). 
The two kernels share the only two features that matter here: they are strongly peaked on the Planckian scale and decay as $e^{-\Omega|t|}$ at large arguments. Consequently, blurring $C$ with either kernel yields the same late-time rate for $F$ and for $T\Psi$. We phrase everything in terms of $F$ below, but identical conclusions hold for $T\Psi$, and for $C$ starting from $\Psi$.

Computing $F$ on the ray $x=avt$ gives
\beq\label{eq:F_saddle_integral}
F(x=avt,t)\;\sim\; \int_{-\infty}^{\infty}\!{\rm d}t'\,
\exp\!\left[-\lambda t'\,\Phi\!\left(\tfrac{x}{v t'}\right)-\,{\cal L}(t-t')\right],
\eeq
where ${\cal L}(s)=-\ln g_\Omega(s)$, with ${\cal L}(s)\simeq\Omega |s|$ for $\Omega|s|\gg 1$. Substituting Eq.~\eqref{eq:LDP_ansatz} and approximating the kernel, the dominant late-time contribution comes from the one-sided integral \footnote{The branch 
$t'<0$ is further suppressed by $e^{-\Omega t}$.}
\beq\label{eq:saddle_integral}
F(x=avt,t)\;\sim\; e^{\Omega t}\,t\!\int_{1}^{\infty}\!{\rm d}\tau\,
e^{\!-\,t\lambda\,\left(\tau\,\Phi(a/\tau)+\frac{\Omega}{\lambda}\tau\right)},
\eeq
after rescaling $t'=t\tau$. The integral is evaluated by saddle point at large $t\lambda$. The saddle condition reads
\beq\label{eq:saddle_condition}
\lambda\,\Phi\!\left(\tfrac{a}{\tau^{\ast}}\right)
-\frac{a\lambda}{\tau^{\ast}}\,\Phi'\!\left(\tfrac{a}{\tau^{\ast}}\right)
+\Omega\;=\;0\ .
\eeq
The saddle dominates when $\tau^{\ast}> 1$; otherwise the integral is controlled by the boundary $\tau=1$. 
This allows us to distinguish two regimes according to the value taken by the FDT ratio $X = \Psi/(\beta C)$ and a generalization $X_{FC} = F/C$ at large times (see Methods): while the first is a signature of classical equilibration, the second does not depend explicitely on $\beta$ and can be viewed as a signature of classicality.
In this way we find:

\textit{Classical regime (the boundary wins).}---When the solution of Eq.~\eqref{eq:saddle_condition} falls at $\tau^{\ast}<1$, the integral is dominated by $\tau=1$, giving
\beq\label{eq:F_classical}
F(x=avt,t)\;\sim\;e^{-\lambda\,t\,\Phi(a)}\;=\;C(x=avt,t)\ .
\eeq
The blurring is ineffective: $F$ inherits the diffusive hydrodynamic decay of $C$, and so does $\Psi$. The three correlators share the same rate and the FDT ratios $X(x,t)\equiv \Psi/(\beta C)\simeq 1$, $X_{FC}\equiv F/C\simeq 1$ are essentially classical. 

\textit{Quantum regime (a genuine saddle).}---Here Eq.~\eqref{eq:saddle_condition} admits a solution $\tau^{\ast} = a f(\Omega/\lambda)$, with $f$ a function whose form depends on $\Phi$,  and $\tau^{\ast}>1$, and the interior saddle controls the integral: $F$ is no longer set by the hydrodynamic rate $\lambda(a)$. One finds
\beq\label{Planckian_F}
F(x=a v t, t) \simeq e^{\Omega t [1 - \frac{x}{v t}  {\mathcal F}(\Omega/\lambda)]} \gg e^{- \lambda t \Phi(\frac{x}{v t})}\ ,
\eeq
where $\mathcal F$ depends on $\Phi$. This is the universal Planckian form $F\sim e^{\Omega t}$ at fixed $x$ near the light cone, very different from the diffusive growth predicted by~\eqref{eq:LDP_ansatz}. The diffusive ansatz therefore generates $X\neq O(1)$ and $X_{FC}\neq O(1)$, signalling the breakdown of classical thermal equilibrium. 

\textit{Outside the causal cone.}---Here Eq.~\eqref{eq:LDP_ansatz} no longer holds for $C$; however, the Lieb–Robinson bound implies $R(x,t)\sim e^{-\mu (x-vt)}$ for $x>vt$, with $\mu$ growing along the cone. The inverse FDT real-time relations (cf.\ Methods) then imply a well known Planckian `leakage' $C\sim e^{2\Omega (t-x/v)}$~\cite{caron2022spacelike}, and the FDT ratio again departs from unity.

Figure~\ref{fig:scheme} summarises this structure. The boundary between the regimes is the set of rays on which the saddle and the integration limit coincide, $\tau^{\ast}=1$. From Eq.~\eqref{eq:saddle_condition} this gives the implicit equation for the critical velocity $a_c$:
\beq\label{eq:ac_def}
a_{c}\,\Phi'(a_{c})-\,\Phi(a_{c})\;=\;\frac{\Omega}{\lambda}\ .
\eeq
Rays with $a<a_c$ have classical diffusive dynamics; rays with $a>a_c$ inherit a Planckian decay rate, since the vicinity of the cone is controlled by inertia rather than friction.

\textbf{Where the Planckian scaling comes from.}
The argument is by contradiction. At low temperature $\beta\to\infty$, $\Omega\to0$. Suppose first that $\lambda$ remains finite, as for a temperature-independent diffusion constant. Then $\Omega/\lambda\to 0$ and the critical ray $a_c$ defined by Eq.~\eqref{eq:ac_def} is forced to zero~\footnote{The left-hand side of~\eqref{eq:ac_def} is an increasing function of $a$ that vanishes at $a=0$ (since $\Phi(0)=0$ and $\Phi'(0)=0$), so that $a_c$ grows monotonically with $\Omega/\lambda$}. The classical region shrinks and one may legitimately expand $\Phi(a)\simeq a^{2}$, recovering the diffusive sector. In that regime Eq.~\eqref{eq:saddle_condition} has the explicit solution
\beq
\tau^{\ast}\;=\;a\,\sqrt{\frac{\lambda}{\Omega}}\ ,
\eeq
which increases with $a$: moving toward the time axis pushes $\tau^{\ast}$ to the boundary $\tau^{\ast}=1$ (classical regime), while moving toward the cone pushes it inside (quantum regime). The critical ray is, combining with~\eqref{eq:lambda_vs_D},
\beq\label{eq:ac_small}
a_c\;\simeq\;\sqrt{\frac{\Omega}{\lambda}}
\;=\;\sqrt{\frac{4D\pi}{v^{2}\hbar \beta}}\ ,
\eeq
the quantitative version of Fig.~\ref{fig:scheme}: the classical wedge is bounded by a ray whose slope depends on the single dimensionless ratio $\Omega/\lambda=4D\Omega/v^{2}$, and which collapses onto the time axis as $T\to0$. Conversely, if one insists that a finite portion of spacetime remain describable by classical hydrodynamics at arbitrarily low temperatures,  $a_c$ must stay finite and, by monotonicity, $\Omega/\lambda$ cannot vanish:
\beq\label{eq:PlBound}
\lambda\;=\;\frac{v^{2}}{4 D}\;\lesssim\;\Omega\;=\;\frac{\pi}{\hbar\beta}\;,
\eeq
a Planckian bound on the dissipation. 

Finally, one can compare our criterion with the bounds derived from analytic properties of the regularised function in Refs.~\cite{maldacena2016bound,qi2026planckian}. For the ansatz~\eqref{eq:LDP_ansatz}, the saddle-point criterion in Eq.~\eqref{eq:ac_def} is equivalent to
\beq\label{Bound_F}
\left|\frac{1}{F(x,t)}\frac{dF(x,t)}{dt}\right|_{x=a_c vt}\;=\;\Omega\ ,
\eeq
i.e.\ the logarithmic decay rate of $F$ on the critical ray saturates the Planckian rate. We thus recover, from a blurring viewpoint, the same threshold found in those works.

\subsection{Phenomenological hydrodynamic equations}\label{Sec_eqs}
To give flesh to the mechanism, we now compute the rate $\Phi(a)$ for three phenomenological equations -- diffusion, telegraph, and diffusive-telegraph -- for a conserved quantity (e.g.\ density or transverse velocity), following the classification of Ref.~\cite{joseph1989heat} (see Appendix~\ref{sec_hydro_theories} for a self-contained summary), where, for instance, such equations are used to model heat transported by attenuated waves. The same structure as the telegraph equation is standard in relativistic fluids, where second-order (Israel--Stewart) hydrodynamics cures the acausality of the Navier--Stokes description by letting dissipative currents relax on a finite time~\cite{israel1976nonstationary,baier2008relativistic,romatschke2010new}. Which of these equations governs a given microscopic quantum system is an open question beyond our scope: here they serve as tractable realisations of the mechanism of Sec.~\ref{Sec_general}. All three equations can be viewed as starting from a linearised hydrodynamics. At this linear level, quantisation is simple: the response and correlation equations decouple and are in fact the same, with different source terms guaranteeing that the FDT conditions hold~\footnote{Calling $G$ the Green function and $S$ the sources:
$G^{-1}*\Psi = S_\Psi \Rightarrow \Psi=G*S_{\Psi}$, $G^{-1}*C = S_C \Rightarrow C=G*S_{C}$. The relation $S_\Psi= D_{\Omega}*S_C$ guarantees the FDT condition $\Psi=D_\Omega*C$.}.

All three constructions start from continuity, $\partial_t n + \nabla j = 0$, and differ only in the constitutive relation between current and gradient. A key feature shared by the telegraph and diffusive-telegraph equations is that the current has a finite relaxation time $\tau_J= (4\lambda)^{-1}$; the Planckian bound $\lambda \lesssim \Omega$ is ultimately a constraint on this relaxation time.

\subsubsection{Diffusion equation}\label{Sec_diff}
Fick's law $j=-D\nabla n$ combined with continuity gives the heat equation $\partial_t n = D\Delta n$, with the familiar Gaussian
\beq\label{eq:Gauss}
C(x,t)\;=\;\frac{1}{\sqrt{4\pi D t}}\;e^{-x^{2}/(4Dt)}\ .
\eeq
Diffusion propagates at infinite speed and has no intrinsic light cone; one must import a maximal velocity $v$ from outside, for instance as a Lieb–Robinson velocity. With this input, the ray parameter $a=x/(vt)$ is meaningful for $a\lesssim 1$ and~\eqref{eq:Gauss} matches the small-$a$ limit of the general ansatz~\eqref{eq:LDP_ansatz}, with $\Phi(a)=a^{2}$ and
\beq
\lambda^{\rm diff}(a)\;=\;\lambda\,a^{2}\ ,\qquad
\lambda\;=\;\frac{v^{2}}{4D}\ .
\eeq
When $v$ is the light-cone velocity, this timescale matches that of Ref.~\cite{hartman2017upper}, which lower-bounds the equilibration time $\tau_{eq}\gtrsim \lambda^{-1}$.

The saddle-point integral~\eqref{eq:saddle_integral} can be evaluated in closed form: $\tau^{\ast}=a\sqrt{\lambda/\Omega}$ and the boundary is crossed at the critical ray $a_c^{\rm diff}=\sqrt{\Omega/\lambda}=\sqrt{4D\Omega/v^{2}}$. The decay of $F$ on the ray has two regimes,
\beq\label{eq:F_diff_cases}
F(x=avt,t)\;\sim\;\frac 1{\sqrt t}
\begin{cases}
\exp\!\left[-\dfrac{a^{2} v^{2}}{4D}\, t\right]
 & \lambda a^{2}<\Omega\ \text{(classical)},\\[2mm]
\exp\!\left[t \Omega \Big( 1 -2 \tfrac{x}{v t}  \sqrt{\tfrac{\lambda}{\Omega}}\Big)\right]
 & \lambda a^{2}>\Omega\ \text{(quantum)},
\end{cases}
\eeq
matching smoothly at $a=a_c^{\rm diff}$. Imposing that the classical window survive the limit $\beta\to\infty$ forces the Planckian scaling~\eqref{eq:PlBound}. Because the light cone here is imposed by hand, we now turn to the telegraph equation.

\subsubsection{Telegraph equation}\label{Sec_telegraph}
If the current relaxes to Fick's law with a finite time $\tau$ (Cattaneo's law $\tau\partial_t j + j = -D\nabla n$), continuity gives the telegraph equation~\cite{joseph1989heat,kolesnik2018slow}
\beq\label{eq:telegraph}
\partial_t^{2}n + 4\lambda \,\partial_t n\;=\;v^{2}\Delta n\ ,
\eeq
with $4\lambda=\tau^{-1}$ and $v^{2}=D/\tau$. Unlike diffusion, the equation is hyperbolic: it interpolates between a wave equation at short times and diffusion at late times, with a sharp light cone $|x|=vt$ built in. A saddle-point evaluation of the Fourier solution at large times gives, for $|a|<1$,
\beq\label{eq:telegraph_ray}
C(x=avt,t)\;\propto\;\frac{1}{\sqrt{t}}\,
\exp\!\left[-2 \lambda \,t\,\bigl(1-\sqrt{1-a^{2}}\bigr)\right]\ ,
\eeq
so that
\beq
\lambda^{\rm tel}(a)\;=\;2 \lambda\,\bigl(1-\sqrt{1-a^{2}}\bigr)\ .
\eeq
At small $a$ this reduces to $\lambda a^{2}$, the diffusive result; at the cone $a\to 1$ it saturates at $\lambda$.

Convolving with the kernel $g_\Omega$ (see Methods) and solving the saddle equation yields $\tau^{\ast}=a(2\lambda+\Omega)/\sqrt{\Omega(4\lambda+\Omega)}$. The boundary $\tau^{\ast}=1$ defines a critical characteristic time
\beq\label{eq:gamma_c_tel}
\lambda^{\rm tel}_c\;= \frac12 \;\frac{\sqrt{1-a^{2}}}{1-\sqrt{1-a^{2}}}\,\Omega\ ,
\eeq
or, at fixed $\lambda$, the critical ray
\beq
\label{ac_tel}
a_c\;=\;\frac{\sqrt{4\lambda/\Omega+1}}{2\lambda/\Omega+1}\ .
\eeq
Below $\lambda_c$, $F$ follows the hydrodynamic decay of $C$; above it, $F$ acquires the Planckian envelope. Collecting the two cases for $\lambda t\gg 1$:
\begin{widetext}
\begin{subnumcases}{F(x=avt,t)  \;\sim\; \frac 1{\sqrt{t}}} \label{eq:F_tel_cases_a}
\exp\!\left[-\lambda^{\rm tel}(a)\, t\right] 
& for $a<a_c$ (classical) \\
\label{eq:F_tel_cases_b}
\exp\!\left[ \Omega t \Big( 1 -\tfrac{x}{v t} \sqrt{\tfrac{\lambda}{\Omega}\big(4+\tfrac{\Omega}{\lambda}\big)}\,\Big)\right]  
& for $a>a_c$ (quantum)
\end{subnumcases}
\end{widetext}
Requiring that the classical window persist at low temperature forces $\lambda/\Omega \lesssim 1$, i.e.\ the same Planckian content as~\eqref{eq:PlBound}. It is worth noting that the classical regime extends to the time axis but \textit{never} reaches the light cone itself: at $a=1$, $\lambda^{\rm tel}_c=0$, so a thin sliver along the cone is always quantum, regardless of how large the dissipation is. This feature reappears numerically below.

\subsubsection{Diffusive telegraph equation}\label{Sec_Diff_tel}
A Jeffreys-type refinement of Cattaneo's law allows the current to respond also to $\nabla\partial_t n$,
$\tau\partial_t j + j = -D\nabla n - \tau\kappa\nabla\partial_t n$, giving
\beq
\partial_t^{2}n + 4\lambda\partial_t n\;=\;v^{2}\Delta n
+ \kappa\,\partial_t\Delta n\ .
\eeq
This still describes a conserved density with an approximate light cone and late-time diffusion, but the crossover is smoothed by the extra gradient term: the light cone is soft rather than sharp, and we only use the behaviour well inside it. The diffusive timescale is again $D=v^2/(4\lambda)$. For  $\epsilon=\sqrt{\kappa \lambda}/v \ll 1$, a small-$k$ saddle of the Fourier solution gives $C(x=avt,t)\propto e^{-\lambda t\,\Phi^{\rm Dtel}(a)}$ with, at small $a$,
\beq\label{eq:Phi_Dtel}
\Phi^{\rm Dtel}(a)\;=\;a^{2}+\Bigl(\frac14-\epsilon^2\Bigr)a^{4}+O(a^{6})\ ,
\eeq
which reduces to the expansion of the telegraph rate function, $a^2+a^4/4$, as $\kappa\to0$. Repeating the blurring analysis, the structure mirrors the telegraph case. In the small-$\Omega$ limit, $a_c=\sqrt{\Omega/\lambda}\,\bigl(1\,-\,(\tfrac{3}{8}-\tfrac32\epsilon^2)\,\Omega/\lambda+\dots\bigr)$; requiring its finiteness at low temperature, the Planckian scaling~\eqref{eq:PlBound} follows identically.

\subsection{Numerical illustration}\label{Sec_num}
We now illustrate the mechanism on the telegraph equation, for which everything can be computed by simple numerical integration. The correlator is obtained from the Fourier representation
\beq\label{eq:Ctel_num}
C(x,t)\;=\;\frac{e^{-2\lambda t}}{2\pi}\int\! dk\,
\cos\!\bigl(\sqrt{(vk)^{2}-(2\lambda)^{2}}\,t\bigr)\,e^{-(bk)^{2}/2}\,\cos(kx)\ ,
\eeq
where $b$ is a short-distance regulator, so that the dynamics is initialised with a localised form and negligible time derivative away from the central peak. We fix $v=1$ and $b=0.4$ throughout.

\begin{figure}[h!]
  \centering \includegraphics[width=0.5\linewidth]{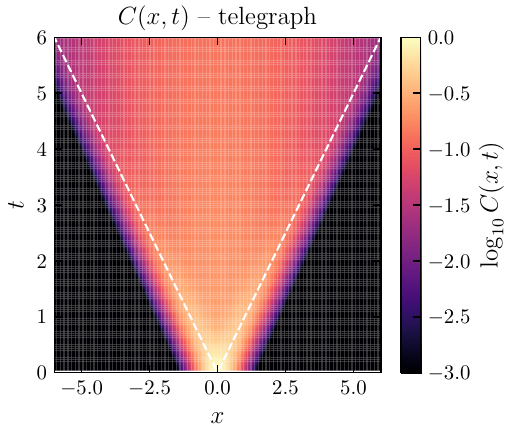}
  \caption{Space-time portrait of the correlator $C(x,t)$ of the telegraph equation~\eqref{eq:telegraph} on a logarithmic colour scale. Dashed white lines mark the light cone $|x|=vt$. Inside the cone the correlator spreads diffusively; outside, it is exponentially suppressed. Parameters: $v=1$, $\lambda=0.35$, $b=0.4$.
  \label{fig:Cxt}}
\end{figure}
\begin{figure}[h!]
  \centering
  \includegraphics[width=0.95\linewidth]{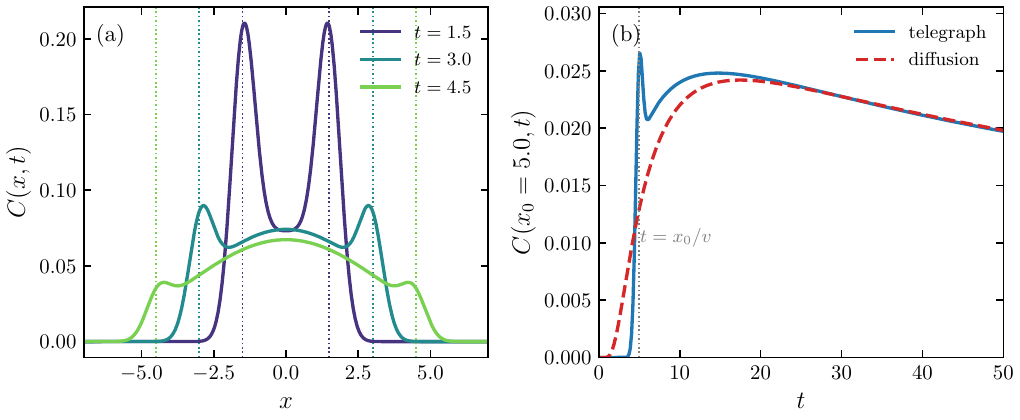}
\caption{Cuts of the correlator. \textbf{(a)} Profile $C(x,t_0)$ at three fixed times, with the corresponding light-cone positions shown as dotted vertical lines. \textbf{(b)} Time evolution $C(x_0,t)$ at fixed position $x_0=5$, compared with the pure-diffusion prediction with $D=v^{2}/(4\lambda)$. The telegraph signal is zero until the light cone arrives at $t=x_0/v$; correlations then rise sharply before settling on the diffusive tail at late times. Parameters as in Fig.~\ref{fig:Cxt}.
  \label{fig:profiles}}
\end{figure}
\begin{figure}[h!]
  \centering
  \includegraphics[width=0.95\linewidth]{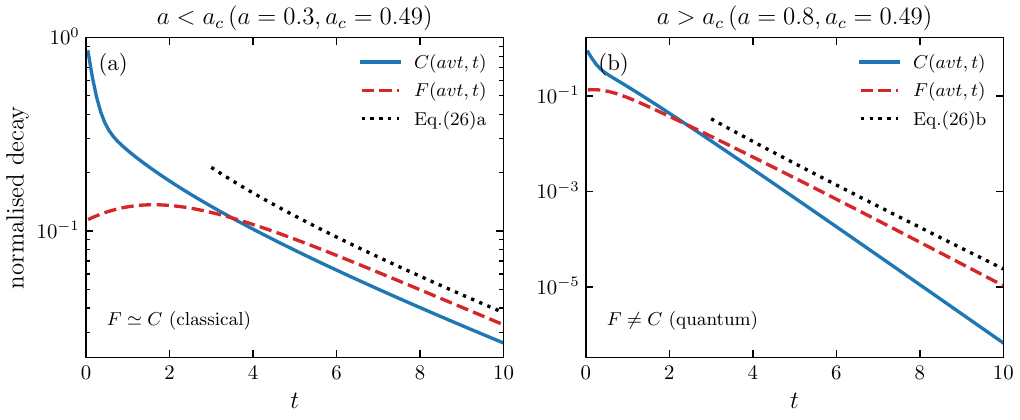}
  \caption{Classical-to-quantum crossover on a ray, at $\lambda=1.75$ and $\beta=6$, as $a$ crosses the critical value $a_c$ from Eq.~\eqref{ac_tel}. We plot $C(avt,t)$ (solid blue) and $F(avt,t)$ (dashed red), obtained by convolving $C$ with $g_\Omega(t)$. The dotted line refers to Eq.\eqref{eq:F_tel_cases_a}. \textbf{(a)} For $a<a_c$ the blurring is trivial, $F\simeq C$, and $X\simeq 1$. \textbf{(b)} For $a>a_c$ the blurring dominates: $F$ tracks the Planckian envelope~\eqref{eq:F_tel_cases_b} (dotted black) rather than $C$.
  \label{fig:Fray}}
\end{figure}

Figure~\ref{fig:Cxt} shows the space-time structure of $C(x,t)$ for $\lambda=0.35$: the expected sharp light cone is clearly visible, with diffusive spreading inside. The two cuts of Fig.~\ref{fig:profiles} make the behaviour at fixed time and fixed position explicit, and show how the telegraph correlator converges to pure diffusion at late times.

Figure~\ref{fig:Fray} is the central numerical result. At $\lambda=1.75$ and $\beta=6$ we compute $F(avt,t)=\int g_\Omega(t-t')C(avt,t')\,dt'$ for two ray slopes on either side of the critical value $a_c=0.49$ from~\eqref{ac_tel}. For $a<a_c$ the blurring kernel is too narrow to alter the decay of $C$ and $F\simeq C$, so the FDT is roughly classical; the corresponding ratios $X_{FC}$ and $X$ are shown in Appendix~\ref{sec_ratios}. For $a>a_c$, $F$ no longer follows $C$ but closely tracks the Planckian exponential~\eqref{eq:F_tel_cases_b}, with the two functions exhibiting markedly different behaviour. This is precisely the classical-to-quantum crossover that drives the Planckian scaling of Sec.~\ref{Sec_general}.

\section{Discussion}\label{Sec_concl}

We have re-examined, from a hydrodynamic point of view, the mechanism by which Planckian timescales arise in quantum many-body systems. The starting point is that the quantum FDT in the time domain amounts to a blurring of the symmetrised correlator on the Planckian scale $\hbar\beta/\pi$. Combining this with generic phenomenological hydrodynamic equations -- diffusion, telegraph, diffusive-telegraph -- in which the correlator has a diffusive large-deviation form $C(x,t)\simeq e^{-\lambda t\Phi(x/(vt))}$ inside a light cone of speed $v$, we find that the interior of the cone divides into two regions. Near the time axis, along a ray $x=avt$, the slowly varying diffusive ansatz makes the blurring trivial: $F$, $\Psi$ and $C$ share the same late-time rate and the system is indistinguishable from a classical one. Closer to the cone the blurring dominates the large-deviation rate function: $F$ and $\Psi$ lose memory of $C$ and behave very differently from the diffusive hydrodynamic form, exhibiting a Planckian growth $\Omega$. The boundary is a critical ray $a_c$ determined by~\eqref{eq:ac_def} and the ratio $\lambda/\Omega$. Requiring  $a_c$ to stay finite as the temperature is lowered bounds the current relaxation rate, $\lambda\lesssim\pi/(\hbar\beta)$, or equivalently $v^2/(4D)\lesssim \pi/(\hbar\beta)$. Within our phenomenological framework, the comparison of the current relaxation time with the Planckian time decides the nature of the dynamics (classical vs.\ quantal) inside the cone.

Written as $DT\gtrsim\hbar v^2/(4\pi k_B)$, the bound only requires that $DT$ does not vanish at low temperature.  This is violated by a temperature-independent $\lambda$, as in a disordered metal. However in this case, the current relaxes by elastic scattering before local equilibrium is reached and condition~\eqref{eq:hydro_assumption} fails.
In fact, in a disordered metal, elastic impurity scattering conserves the energy of each particle, so that every energy shell diffuses independently and no local temperature is established; the current relaxes on the elastic time $\tau_J\sim\tau_{\rm el}$, well before local equilibration, which is due to inelastic processes, $\tau_{\rm hydro}\sim\tau_{\rm in}$, slow at low temperature. The resulting diffusion constant $D \propto v_F^2\tau_{\rm el}$ is perfectly well defined, but $v_F$ has no hydrodynamic meaning and such single-particle diffusion falls outside our framework and
the vanishing of the classical region is no contradiction here. More generally, for $\lambda=\lambda_0+\alpha k_BT/\hbar$ the criterion applies only when inelastic processes dominate current relaxation, so that the bound concerns the inelastic part of the rate, as emphasised in Ref.~\cite{hartnoll2022colloquium}.

Our derivation is generic and it does not depend on the underlying microscopic mode, but it is conditional. The bound follows from demanding that a classical hydrodynamic description be internally consistent with the quantum FDT down to low temperatures. Planckian timescales are thus not a mysterious quantum manifestation: they are the price one pays to keep describing a quantum system with classical hydrodynamics up to the lowest temperature, provided transport is genuinely hydrodynamic, $\tau_{\rm hydro}\leq\tau_J$.

It is worth commenting on the relation to Ref.~\cite{qi2026planckian}, which uses analyticity properties of the regulated correlator $F$ to derive a Planckian lower bound on the early timescale after which the algebraic decay of hydrodynamics sets in -- a local equilibration time. On the other hand, we work at late times along rays, assuming the existence of a light-cone, such that
 the correlator decays exponentially. Our classicality criterion [cf.\ Eq.~\eqref{eq:ac_def}] amounts to requiring that $F, C, \Psi/\beta$ share the same decay rate and assuming its validity down to low temperature, one concludes a bound on the diffusion constant.

Several directions seem worth pursuing. First, our argument leaves open the fate of systems that violate~\eqref{eq:PlBound}: the classical hydrodynamic description encoded in our starting point becomes self-inconsistent with the quantum FDT, and one should presumably switch to a genuinely quantum hydrodynamics to describe the approach to the cone.
Second, our bound can be related to other conjectured Planckian bounds on transport, provided the same physical velocity $v$ can be identified in the different relations.

The scaling~\eqref{eq:PlBound}, rewritten as $D\gtrsim v^{2}\hbar\beta/(4\pi)$, then lower-bounds the local equilibration time of Ref.~\cite{hartman2017upper}: since $\tau_{\rm eq}\gtrsim D/v^{2}$~\cite{hartman2017upper,qi2026planckian,delacretaz2025bound}, our inequality implies $\tau_{\rm eq}\gtrsim\hbar\beta$. Moreover, identifying $D$ with the momentum diffusivity $D_{\eta}=\eta c^2/(s T)$ of a relativistic fluid~\cite{hartnoll2022colloquium}, setting $v=c$, and assuming that the shear channel obeys an Israel--Stewart (telegraph-type) equation, the bound $D\gtrsim v^{2}\hbar\beta/(4\pi)$ translates into $\eta/s\gtrsim \hbar/(4\pi k_{B})$, which has the form of the Kovtun–Son–Starinets bound~\cite{kovtun2005viscosity}, with a prefactor that our argument does not fix. It would be interesting to make these connections more precise, and to understand whether the pattern of classical and quantum regions we find near the cone can shed light on the tightness of these inequalities. Finally, an explicit microscopic derivation or observation of the different classical and quantum regimes in an interacting quantum model would be a rewarding next step; a natural candidate is the SYK chain, where in the large-$q$ limit the energy-density correlators, the diffusion constant and the butterfly velocity are known in closed form at all couplings~\cite{choi2021pole}.

\section*{Methods}\label{Sec_def}
\subsection*{Fluctuation-dissipation theorem and its ratio}

We consider a quantum system at inverse temperature $\beta$, governed by a Hamiltonian $H$, and a space-dependent Hermitian observable $A(x,t)$. The information about two-point dynamical correlations is encoded in four related functions (we assume $\langle A\rangle=0$ throughout):
\begin{subequations}
\label{eq:CRdef}
\begin{align}
\text{symmetrised correlator:}\quad \displaystyle
    C(x,t) &\displaystyle = \frac{1}{2Z}\,
    \mathrm{Tr}\,\Big[e^{-\beta H}\{A(x,t),A(0,0)\}\Big]
    \\[5pt]
\text{regulated correlator:}\quad
    \displaystyle F(x,t) &\displaystyle = \frac{1}{Z}\,
    \mathrm{Tr}\,\Big[e^{-\tfrac{\beta}{2}H}A(x,t)\,e^{-\tfrac{\beta}{2}H}A(0,0)\Big]
       \\[5pt]
\text{response:}\quad
   \displaystyle R(x,t) &\displaystyle = \frac{i}{\hbar}\theta(t)\frac{1}{Z}\,
    \mathrm{Tr}\,\Big[e^{-\beta H}[A(x,t),A(0,0)]\Big]
         \\[5pt]
\label{eq:Psi_def}
\text{integrated response:}\quad
   \displaystyle \Psi(x,t) &\displaystyle = \int_{|t|}^{\infty}\!\!R(x,t')\,dt'.
\end{align}
\end{subequations}


These four functions are related by the fluctuation-dissipation theorem. While $C$ and $R$ are the standard objects of linear response, $F$ -- in which the two operators are separated by half a thermal circle -- is the one whose analytic properties underlie the bound on chaos~\cite{maldacena2016bound} and the bound on the local equilibration time~\cite{qi2026planckian}. As recalled below, it is obtained from $C$ by a blurring on the Planckian time scale and coincides with it in the classical limit; since $\Psi$ has a more direct physical meaning, we follow it in parallel.

\paragraph*{The quantum FDT as a blurring.}
The quantum FDT takes a simple form in frequency~\cite{mahan2000many}, but it is much more illuminating in the time domain~\cite{pottier2001quantum,pappalardi2021quantum}, where it can be recast as two convolutions:
\begin{subequations}\label{eq:blurring}
\begin{align}
\label{eq:FoConvC}
F(x,t) &= \int_{-\infty}^{\infty}
g_{\Omega}(t-t')\,C(x,t')\,dt',\\
\label{eq:PsioConvC}
\Psi(x,t) &= \beta \int_{-\infty}^{\infty}
D_{\Omega}(t-t')\,C(x,t')\,dt',
\end{align}
\end{subequations}
with the first Matsubara frequency $\Omega$ defined in Eq.~\eqref{eq_defOmega} and kernels
\beq
g_{\Omega}(t) \;=\; \frac{\Omega}{\pi}\,\frac{1}{\cosh(\Omega t)}\ ,\quad
D_{\Omega}(t) \;=\; -\frac{2\Omega}{\pi^{2}}\,
\ln\tanh\!\left(\frac{\Omega|t|}{2}\right) \label{blurr12}
\eeq
that are positive, normalised to one, strongly peaked on the Planckian scale $\Omega^{-1}=\hbar\beta/\pi$, and decay as $e^{-\Omega|t|}$ at large times. The passage from $\Psi(x,t)$ to $C(x,t)$ is less simple but can also be written as a series of convolutions (see Appendix~\ref{Sec_dyadic}):
\begin{subequations}
\begin{align}\label{eq:C_dyadic1}
\beta\,C(x,t) & -\Psi(x,t) \nonumber 
=
\sum_{k=0}^{\infty}2^k\!
\left[
\Psi(x,t)
-
\int_{-\infty}^{\infty}dt'\,G_{2^k\Omega}(t-t')\,\Psi(x,t')
\right]
\end{align}
with
\begin{equation}\label{eq:FT1}
G_\Omega(t)
=
\frac{\Omega}{2}\,
\frac{1}{\cosh^2\!\!\left(\Omega t\right)}
\end{equation}
(the $G_{2^k \Omega}$ become more peaked, so terms get smaller with $k$; see Appendix~\ref{Sec_dyadic}).
\end{subequations}

Passing from $C$ to $F$ or $\Psi$ amounts to \textit{blurring} $C$ on the Planckian timescale: features finer than $\hbar\beta/\pi$ are washed out, while slower features are essentially untouched. In the classical limit $\hbar\beta\to 0$, $\Omega\to\infty$ and the kernels collapse to Dirac deltas, so $F(x,t)=C(x,t)$ and $T\Psi(x,t)=C(x,t)$.

\paragraph*{The fluctuation-dissipation ratio.}
A useful diagnostic of classical equilibrium is the fluctuation-dissipation ratio~\cite{cugliandolo1997energy}:
\beq\label{eq:FDT_ratio}
X(x, t)\;\equiv\;\frac{\Psi(x,t)}{\beta\, C(x,t)} \ .
\eeq
When $X(x,t)\simeq 1$ the dynamics can be interpreted as that of a classical system in thermal equilibrium; deviations signal an out-of-equilibrium regime and sometimes allow for the definition of effective temperatures. In our context we are in equilibrium, so strong deviations from unity instead indicate that the diffusive hydrodynamic solution from which we start is inconsistent with classical equilibration. We also define
\beq\label{eq:FDT_ratio_FC}
X_{FC}(x, t)\;\equiv\;\frac{F(x,t)}{C(x,t)}\ ,
\eeq
in which temperature does not appear explicitly; $X_{FC}\simeq 1$ is again a criterion of classicality, since $F$ and $C$ coincide in the classical limit. The two ratios identify the same classical region and the same critical ray (Appendix~\ref{sec_ratios}).

\medskip
\acknowledgments
We thank X. Cao and L. Delacr\'etaz for important discussions. L.F. thanks J.-Y. Ollitrault, M. Winn, and F. Gelis for clarifications around the viscosity bound. S.P. thanks X. Turkeshi for useful suggestions. This work was supported by the French government through the France 2030 program (PhOM -- Graduate School of Physics), under reference ANR-11-IDEX-0003 (Project Mascotte, L. Foini). S.P. acknowledges funding by the Deutsche Forschungsgemeinschaft (DFG, German Research Foundation) under Germany's Excellence Strategy -- Cluster of Excellence Matter and Light for Quantum Computing (ML4Q) EXC 2004/1 -- 390534769 and the DFG Collaborative Research Center (CRC) 183 Project No.\ 277101999 -- project B02.

\medskip
\section*{Author contributions}
All authors have contributed equally to the ideation, realization and writing of the paper.

\medskip
\section*{Code Availability}
The code generated in this study have been deposited in the Zenodo
public folder \cite{zenodo}.

\bibliography{biblio}

\begin{thebibliography}{47}%
\makeatletter
\providecommand \@ifxundefined [1]{%
 \@ifx{#1\undefined}
}%
\providecommand \@ifnum [1]{%
 \ifnum #1\expandafter \@firstoftwo
 \else \expandafter \@secondoftwo
 \fi
}%
\providecommand \@ifx [1]{%
 \ifx #1\expandafter \@firstoftwo
 \else \expandafter \@secondoftwo
 \fi
}%
\providecommand \natexlab [1]{#1}%
\providecommand \enquote  [1]{``#1''}%
\providecommand \bibnamefont  [1]{#1}%
\providecommand \bibfnamefont [1]{#1}%
\providecommand \citenamefont [1]{#1}%
\providecommand \href@noop [0]{\@secondoftwo}%
\providecommand \href [0]{\begingroup \@sanitize@url \@href}%
\providecommand \@href[1]{\@@startlink{#1}\@@href}%
\providecommand \@@href[1]{\endgroup#1\@@endlink}%
\providecommand \@sanitize@url [0]{\catcode `\\12\catcode `\$12\catcode
  `\&12\catcode `\#12\catcode `\^12\catcode `\_12\catcode `\%12\relax}%
\providecommand \@@startlink[1]{}%
\providecommand \@@endlink[0]{}%
\providecommand \url  [0]{\begingroup\@sanitize@url \@url }%
\providecommand \@url [1]{\endgroup\@href {#1}{\urlprefix }}%
\providecommand \urlprefix  [0]{URL }%
\providecommand \Eprint [0]{\href }%
\providecommand \doibase [0]{https://doi.org/}%
\providecommand \selectlanguage [0]{\@gobble}%
\providecommand \bibinfo  [0]{\@secondoftwo}%
\providecommand \bibfield  [0]{\@secondoftwo}%
\providecommand \translation [1]{[#1]}%
\providecommand \BibitemOpen [0]{}%
\providecommand \bibitemStop [0]{}%
\providecommand \bibitemNoStop [0]{.\EOS\space}%
\providecommand \EOS [0]{\spacefactor3000\relax}%
\providecommand \BibitemShut  [1]{\csname bibitem#1\endcsname}%
\let\auto@bib@innerbib\@empty
\bibitem [{\citenamefont {Hartnoll}\ and\ \citenamefont
  {Mackenzie}(2022)}]{hartnoll2022colloquium}%
  \BibitemOpen
  \bibfield  {author} {\bibinfo {author} {\bibfnamefont {S.~A.}\ \bibnamefont
  {Hartnoll}}\ and\ \bibinfo {author} {\bibfnamefont {A.~P.}\ \bibnamefont
  {Mackenzie}},\ }\bibfield  {title} {\bibinfo {title} {Colloquium: Planckian
  dissipation in metals},\ }\href
  {https://doi.org/10.1103/RevModPhys.94.041002} {\bibfield  {journal}
  {\bibinfo  {journal} {Rev. of Mod. Phys.}\ }\textbf {\bibinfo {volume}
  {94}},\ \bibinfo {pages} {041002} (\bibinfo {year} {2022})}\BibitemShut
  {NoStop}%
\bibitem [{\citenamefont {Chowdhury}\ \emph {et~al.}(2022)\citenamefont
  {Chowdhury}, \citenamefont {Georges}, \citenamefont {Parcollet},\ and\
  \citenamefont {Sachdev}}]{chowdhury2022sachdev}%
  \BibitemOpen
  \bibfield  {author} {\bibinfo {author} {\bibfnamefont {D.}~\bibnamefont
  {Chowdhury}}, \bibinfo {author} {\bibfnamefont {A.}~\bibnamefont {Georges}},
  \bibinfo {author} {\bibfnamefont {O.}~\bibnamefont {Parcollet}},\ and\
  \bibinfo {author} {\bibfnamefont {S.}~\bibnamefont {Sachdev}},\ }\bibfield
  {title} {\bibinfo {title} {Sachdev-ye-kitaev models and beyond: Window into
  non-fermi liquids},\ }\href {https://doi.org/10.1103/RevModPhys.94.035004}
  {\bibfield  {journal} {\bibinfo  {journal} {Rev. of Mod. Phys.}\ }\textbf
  {\bibinfo {volume} {94}},\ \bibinfo {pages} {035004} (\bibinfo {year}
  {2022})}\BibitemShut {NoStop}%
\bibitem [{\citenamefont {Zaanen}(2004)}]{zaanen2004temperature}%
  \BibitemOpen
  \bibfield  {author} {\bibinfo {author} {\bibfnamefont {J.}~\bibnamefont
  {Zaanen}},\ }\bibfield  {title} {\bibinfo {title} {Why the temperature is
  high},\ }\href {https://doi.org/10.1038/430512a} {\bibfield  {journal}
  {\bibinfo  {journal} {Nature}\ }\textbf {\bibinfo {volume} {430}},\ \bibinfo
  {pages} {512} (\bibinfo {year} {2004})}\BibitemShut {NoStop}%
\bibitem [{\citenamefont {Kovtun}\ \emph {et~al.}(2005)\citenamefont {Kovtun},
  \citenamefont {Son},\ and\ \citenamefont {Starinets}}]{kovtun2005viscosity}%
  \BibitemOpen
  \bibfield  {author} {\bibinfo {author} {\bibfnamefont {P.}~\bibnamefont
  {Kovtun}}, \bibinfo {author} {\bibfnamefont {D.~T.}\ \bibnamefont {Son}},\
  and\ \bibinfo {author} {\bibfnamefont {A.~O.}\ \bibnamefont {Starinets}},\
  }\bibfield  {title} {\bibinfo {title} {Viscosity in strongly interacting
  quantum field theories from black hole physics},\ }\href
  {https://doi.org/10.1103/PhysRevLett.94.111601} {\bibfield  {journal}
  {\bibinfo  {journal} {Phys. Rev. Lett.}\ }\textbf {\bibinfo {volume} {94}},\
  \bibinfo {pages} {111601} (\bibinfo {year} {2005})}\BibitemShut {NoStop}%
\bibitem [{\citenamefont {Maldacena}\ \emph {et~al.}(2016)\citenamefont
  {Maldacena}, \citenamefont {Shenker},\ and\ \citenamefont
  {Stanford}}]{maldacena2016bound}%
  \BibitemOpen
  \bibfield  {author} {\bibinfo {author} {\bibfnamefont {J.}~\bibnamefont
  {Maldacena}}, \bibinfo {author} {\bibfnamefont {S.~H.}\ \bibnamefont
  {Shenker}},\ and\ \bibinfo {author} {\bibfnamefont {D.}~\bibnamefont
  {Stanford}},\ }\bibfield  {title} {\bibinfo {title} {A bound on chaos},\
  }\href {https://doi.org/10.1007/JHEP08(2016)106} {\bibfield  {journal}
  {\bibinfo  {journal} {Journal of High Energy Physics}\ }\textbf {\bibinfo
  {volume} {2016}},\ \bibinfo {pages} {106} (\bibinfo {year}
  {2016})}\BibitemShut {NoStop}%
\bibitem [{\citenamefont {Hartman}\ \emph {et~al.}(2017)\citenamefont
  {Hartman}, \citenamefont {Hartnoll},\ and\ \citenamefont
  {Mahajan}}]{hartman2017upper}%
  \BibitemOpen
  \bibfield  {author} {\bibinfo {author} {\bibfnamefont {T.}~\bibnamefont
  {Hartman}}, \bibinfo {author} {\bibfnamefont {S.~A.}\ \bibnamefont
  {Hartnoll}},\ and\ \bibinfo {author} {\bibfnamefont {R.}~\bibnamefont
  {Mahajan}},\ }\bibfield  {title} {\bibinfo {title} {Upper bound on
  diffusivity},\ }\href {https://doi.org/10.1103/PhysRevLett.119.141601}
  {\bibfield  {journal} {\bibinfo  {journal} {Phys. Rev. Lett.}\ }\textbf
  {\bibinfo {volume} {119}},\ \bibinfo {pages} {141601} (\bibinfo {year}
  {2017})}\BibitemShut {NoStop}%
\bibitem [{\citenamefont {Delacretaz}(2025)}]{delacretaz2025bound}%
  \BibitemOpen
  \bibfield  {author} {\bibinfo {author} {\bibfnamefont {L.~V.}\ \bibnamefont
  {Delacretaz}},\ }\bibfield  {title} {\bibinfo {title} {A bound on
  thermalization from diffusive fluctuations},\ }\href
  {https://doi.org/10.1038/s41567-025-02810-2} {\bibfield  {journal} {\bibinfo
  {journal} {Nature Physics}\ }\textbf {\bibinfo {volume} {21}},\ \bibinfo
  {pages} {669} (\bibinfo {year} {2025})}\BibitemShut {NoStop}%
\bibitem [{\citenamefont {Qi}\ \emph {et~al.}(2026)\citenamefont {Qi},
  \citenamefont {Milekhin},\ and\ \citenamefont
  {Delacr{\'e}taz}}]{qi2026planckian}%
  \BibitemOpen
  \bibfield  {author} {\bibinfo {author} {\bibfnamefont {M.}~\bibnamefont
  {Qi}}, \bibinfo {author} {\bibfnamefont {A.}~\bibnamefont {Milekhin}},\ and\
  \bibinfo {author} {\bibfnamefont {L.}~\bibnamefont {Delacr{\'e}taz}},\
  }\bibfield  {title} {\bibinfo {title} {Planckian bound on the local
  equilibration time},\ }\href {https://arxiv.org/abs/2602.17638} {\bibfield
  {journal} {\bibinfo  {journal} {arXiv preprint arXiv:2602.17638}\ } (\bibinfo
  {year} {2026})}\BibitemShut {NoStop}%
\bibitem [{\citenamefont {Cao}(2025)}]{cao2025planckian}%
  \BibitemOpen
  \bibfield  {author} {\bibinfo {author} {\bibfnamefont {X.}~\bibnamefont
  {Cao}},\ }\bibfield  {title} {\bibinfo {title} {Planckian bound on quantum
  dynamical entropy},\ }\href {https://arxiv.org/abs/2507.20914} {\bibfield
  {journal} {\bibinfo  {journal} {arXiv preprint arXiv:2507.20914}\ } (\bibinfo
  {year} {2025})}\BibitemShut {NoStop}%
\bibitem [{\citenamefont {Nussinov}\ and\ \citenamefont
  {Chakrabarty}(2022)}]{nussinov2022exact}%
  \BibitemOpen
  \bibfield  {author} {\bibinfo {author} {\bibfnamefont {Z.}~\bibnamefont
  {Nussinov}}\ and\ \bibinfo {author} {\bibfnamefont {S.}~\bibnamefont
  {Chakrabarty}},\ }\bibfield  {title} {\bibinfo {title} {Exact universal
  chaos, speed limit, acceleration, planckian transport coefficient,
  “collapse” to equilibrium, and other bounds in thermal quantum systems},\
  }\href {https://doi.org/10.1016/j.aop.2022.168970} {\bibfield  {journal}
  {\bibinfo  {journal} {Annals of Physics}\ }\textbf {\bibinfo {volume}
  {443}},\ \bibinfo {pages} {168970} (\bibinfo {year} {2022})}\BibitemShut
  {NoStop}%
\bibitem [{\citenamefont {Nussinov}\ and\ \citenamefont
  {Chakrabarty}(2025)}]{nussinov2025planckian}%
  \BibitemOpen
  \bibfield  {author} {\bibinfo {author} {\bibfnamefont {Z.}~\bibnamefont
  {Nussinov}}\ and\ \bibinfo {author} {\bibfnamefont {S.}~\bibnamefont
  {Chakrabarty}},\ }\bibfield  {title} {\bibinfo {title} {Planckian bounds from
  local uncertainty relations},\ }\href
  {https://doi.org/10.1016/j.physc.2025.1354755} {\bibfield  {journal}
  {\bibinfo  {journal} {Physica C: Superconductivity and its Applications}\
  }\textbf {\bibinfo {volume} {636}},\ \bibinfo {pages} {1354755} (\bibinfo
  {year} {2025})}\BibitemShut {NoStop}%
\bibitem [{\citenamefont {Volovik}(2022)}]{volovik2022quantum}%
  \BibitemOpen
  \bibfield  {author} {\bibinfo {author} {\bibfnamefont {G.~E.}\ \bibnamefont
  {Volovik}},\ }\bibfield  {title} {\bibinfo {title} {Quantum turbulence and
  planckian dissipation},\ }\href {https://doi.org/10.1134/s0021364022100344}
  {\bibfield  {journal} {\bibinfo  {journal} {JETP Letters}\ }\textbf {\bibinfo
  {volume} {115}},\ \bibinfo {pages} {461–465} (\bibinfo {year}
  {2022})}\BibitemShut {NoStop}%
\bibitem [{\citenamefont {Hartnoll}(2015)}]{hartnoll2015theory}%
  \BibitemOpen
  \bibfield  {author} {\bibinfo {author} {\bibfnamefont {S.~A.}\ \bibnamefont
  {Hartnoll}},\ }\bibfield  {title} {\bibinfo {title} {Theory of universal
  incoherent metallic transport},\ }\href {https://doi.org/10.1038/nphys3174}
  {\bibfield  {journal} {\bibinfo  {journal} {Nature Physics}\ }\textbf
  {\bibinfo {volume} {11}},\ \bibinfo {pages} {54} (\bibinfo {year}
  {2015})}\BibitemShut {NoStop}%
\bibitem [{\citenamefont {Aldape}\ \emph {et~al.}(2022)\citenamefont {Aldape},
  \citenamefont {Cookmeyer}, \citenamefont {Patel},\ and\ \citenamefont
  {Altman}}]{aldape2022solvable}%
  \BibitemOpen
  \bibfield  {author} {\bibinfo {author} {\bibfnamefont {E.~E.}\ \bibnamefont
  {Aldape}}, \bibinfo {author} {\bibfnamefont {T.}~\bibnamefont {Cookmeyer}},
  \bibinfo {author} {\bibfnamefont {A.~A.}\ \bibnamefont {Patel}},\ and\
  \bibinfo {author} {\bibfnamefont {E.}~\bibnamefont {Altman}},\ }\bibfield
  {title} {\bibinfo {title} {Solvable theory of a strange metal at the
  breakdown of a heavy {F}ermi liquid},\ }\href
  {https://doi.org/10.1103/PhysRevB.105.235111} {\bibfield  {journal} {\bibinfo
   {journal} {Phys. Rev. B}\ }\textbf {\bibinfo {volume} {105}},\ \bibinfo
  {pages} {235111} (\bibinfo {year} {2022})}\BibitemShut {NoStop}%
\bibitem [{\citenamefont {Patel}\ \emph {et~al.}(2023)\citenamefont {Patel},
  \citenamefont {Guo}, \citenamefont {Esterlis},\ and\ \citenamefont
  {Sachdev}}]{patel2023universal}%
  \BibitemOpen
  \bibfield  {author} {\bibinfo {author} {\bibfnamefont {A.~A.}\ \bibnamefont
  {Patel}}, \bibinfo {author} {\bibfnamefont {H.}~\bibnamefont {Guo}}, \bibinfo
  {author} {\bibfnamefont {I.}~\bibnamefont {Esterlis}},\ and\ \bibinfo
  {author} {\bibfnamefont {S.}~\bibnamefont {Sachdev}},\ }\bibfield  {title}
  {\bibinfo {title} {Universal theory of strange metals from spatially random
  interactions},\ }\href {https://doi.org/10.1126/science.abq6011} {\bibfield
  {journal} {\bibinfo  {journal} {Science}\ }\textbf {\bibinfo {volume}
  {381}},\ \bibinfo {pages} {790} (\bibinfo {year} {2023})}\BibitemShut
  {NoStop}%
\bibitem [{\citenamefont {Patel}\ \emph {et~al.}(2025)\citenamefont {Patel},
  \citenamefont {Lunts},\ and\ \citenamefont {Albergo}}]{patel2025strange}%
  \BibitemOpen
  \bibfield  {author} {\bibinfo {author} {\bibfnamefont {A.~A.}\ \bibnamefont
  {Patel}}, \bibinfo {author} {\bibfnamefont {P.}~\bibnamefont {Lunts}},\ and\
  \bibinfo {author} {\bibfnamefont {M.~S.}\ \bibnamefont {Albergo}},\
  }\bibfield  {title} {\bibinfo {title} {Strange metals and {P}lanckian
  transport in a gapless phase from spatially random interactions},\ }\href
  {https://doi.org/10.1103/PhysRevX.15.031064} {\bibfield  {journal} {\bibinfo
  {journal} {Phys. Rev. X}\ }\textbf {\bibinfo {volume} {15}},\ \bibinfo
  {pages} {031064} (\bibinfo {year} {2025})}\BibitemShut {NoStop}%
\bibitem [{\citenamefont {Ciuchi}\ and\ \citenamefont
  {Fratini}(2023)}]{ciuchi2023strange}%
  \BibitemOpen
  \bibfield  {author} {\bibinfo {author} {\bibfnamefont {S.}~\bibnamefont
  {Ciuchi}}\ and\ \bibinfo {author} {\bibfnamefont {S.}~\bibnamefont
  {Fratini}},\ }\bibfield  {title} {\bibinfo {title} {Strange metal behavior
  from incoherent carriers scattered by local moments},\ }\href
  {https://doi.org/10.1103/PhysRevB.108.235173} {\bibfield  {journal} {\bibinfo
   {journal} {Phys. Rev. B}\ }\textbf {\bibinfo {volume} {108}},\ \bibinfo
  {pages} {235173} (\bibinfo {year} {2023})}\BibitemShut {NoStop}%
\bibitem [{\citenamefont {Lee}\ and\ \citenamefont
  {Murthy}(2026)}]{lee2026apparent}%
  \BibitemOpen
  \bibfield  {author} {\bibinfo {author} {\bibfnamefont {B.~Y.-H.}\
  \bibnamefont {Lee}}\ and\ \bibinfo {author} {\bibfnamefont {C.}~\bibnamefont
  {Murthy}},\ }\bibfield  {title} {\bibinfo {title} {Apparent {P}lanckian
  scattering from local polaron formation},\ }\href
  {https://arxiv.org/abs/2604.22029} {\bibfield  {journal} {\bibinfo  {journal}
  {arXiv preprint arXiv:2604.22029}\ } (\bibinfo {year} {2026})}\BibitemShut
  {NoStop}%
\bibitem [{\citenamefont {Murthy}\ \emph {et~al.}(2023)\citenamefont {Murthy},
  \citenamefont {Pandey}, \citenamefont {Esterlis},\ and\ \citenamefont
  {Kivelson}}]{murthy2023stability}%
  \BibitemOpen
  \bibfield  {author} {\bibinfo {author} {\bibfnamefont {C.}~\bibnamefont
  {Murthy}}, \bibinfo {author} {\bibfnamefont {A.}~\bibnamefont {Pandey}},
  \bibinfo {author} {\bibfnamefont {I.}~\bibnamefont {Esterlis}},\ and\
  \bibinfo {author} {\bibfnamefont {S.~A.}\ \bibnamefont {Kivelson}},\
  }\bibfield  {title} {\bibinfo {title} {A stability bound on the {T}-linear
  resistivity of conventional metals},\ }\href
  {https://doi.org/10.1073/pnas.2216241120} {\bibfield  {journal} {\bibinfo
  {journal} {Proc. Natl. Acad. Sci. U.S.A.}\ }\textbf {\bibinfo {volume}
  {120}},\ \bibinfo {pages} {e2216241120} (\bibinfo {year} {2023})}\BibitemShut
  {NoStop}%
\bibitem [{\citenamefont {Chowdhury}(2026)}]{chowdhury2026information}%
  \BibitemOpen
  \bibfield  {author} {\bibinfo {author} {\bibfnamefont {D.}~\bibnamefont
  {Chowdhury}},\ }\bibfield  {title} {\bibinfo {title} {Information,
  dissipation, and {P}lanckian optimality},\ }\href
  {https://arxiv.org/abs/2602.04953} {\bibfield  {journal} {\bibinfo  {journal}
  {arXiv preprint arXiv:2602.04953}\ } (\bibinfo {year} {2026})}\BibitemShut
  {NoStop}%
\bibitem [{\citenamefont {Aydin}\ \emph {et~al.}(2024)\citenamefont {Aydin},
  \citenamefont {Keski-Rahkonen},\ and\ \citenamefont
  {Heller}}]{aydin2024quantum}%
  \BibitemOpen
  \bibfield  {author} {\bibinfo {author} {\bibfnamefont {A.}~\bibnamefont
  {Aydin}}, \bibinfo {author} {\bibfnamefont {J.}~\bibnamefont
  {Keski-Rahkonen}},\ and\ \bibinfo {author} {\bibfnamefont {E.~J.}\
  \bibnamefont {Heller}},\ }\bibfield  {title} {\bibinfo {title} {Quantum
  acoustics unravels {P}lanckian resistivity},\ }\href
  {https://doi.org/10.1073/pnas.2404853121} {\bibfield  {journal} {\bibinfo
  {journal} {Proc. Natl. Acad. Sci. U.S.A.}\ }\textbf {\bibinfo {volume}
  {121}},\ \bibinfo {pages} {e2404853121} (\bibinfo {year} {2024})}\BibitemShut
  {NoStop}%
\bibitem [{\citenamefont {Mezei}\ and\ \citenamefont
  {S{\'a}rosi}(2020)}]{mezei2020chaos}%
  \BibitemOpen
  \bibfield  {author} {\bibinfo {author} {\bibfnamefont {M.}~\bibnamefont
  {Mezei}}\ and\ \bibinfo {author} {\bibfnamefont {G.}~\bibnamefont
  {S{\'a}rosi}},\ }\bibfield  {title} {\bibinfo {title} {Chaos in the butterfly
  cone},\ }\href {https://doi.org/10.1007/JHEP01(2020)186} {\bibfield
  {journal} {\bibinfo  {journal} {Journal of High Energy Physics}\ }\textbf
  {\bibinfo {volume} {2020}},\ \bibinfo {pages} {186} (\bibinfo {year}
  {2020})}\BibitemShut {NoStop}%
\bibitem [{\citenamefont {Tsuji}\ \emph {et~al.}(2018)\citenamefont {Tsuji},
  \citenamefont {Shitara},\ and\ \citenamefont {Ueda}}]{tsuji2018bound}%
  \BibitemOpen
  \bibfield  {author} {\bibinfo {author} {\bibfnamefont {N.}~\bibnamefont
  {Tsuji}}, \bibinfo {author} {\bibfnamefont {T.}~\bibnamefont {Shitara}},\
  and\ \bibinfo {author} {\bibfnamefont {M.}~\bibnamefont {Ueda}},\ }\bibfield
  {title} {\bibinfo {title} {Bound on the exponential growth rate of
  out-of-time-ordered correlators},\ }\href
  {https://doi.org/10.1103/physreve.98.012216} {\bibfield  {journal} {\bibinfo
  {journal} {Physical Review E}\ }\textbf {\bibinfo {volume} {98}} (\bibinfo
  {year} {2018})}\BibitemShut {NoStop}%
\bibitem [{\citenamefont {Pappalardi}\ \emph {et~al.}(2022)\citenamefont
  {Pappalardi}, \citenamefont {Foini},\ and\ \citenamefont
  {Kurchan}}]{pappalardi2021quantum}%
  \BibitemOpen
  \bibfield  {author} {\bibinfo {author} {\bibfnamefont {S.}~\bibnamefont
  {Pappalardi}}, \bibinfo {author} {\bibfnamefont {L.}~\bibnamefont {Foini}},\
  and\ \bibinfo {author} {\bibfnamefont {J.}~\bibnamefont {Kurchan}},\
  }\bibfield  {title} {\bibinfo {title} {Quantum bounds and
  fluctuation-dissipation relations},\ }\href
  {https://doi.org/10.21468/SciPostPhys.12.4.130} {\bibfield  {journal}
  {\bibinfo  {journal} {SciPost Physics}\ }\textbf {\bibinfo {volume} {12}},\
  \bibinfo {pages} {130} (\bibinfo {year} {2022})}\BibitemShut {NoStop}%
\bibitem [{\citenamefont {Lieb}\ and\ \citenamefont
  {Robinson}(1972)}]{lieb1972finite}%
  \BibitemOpen
  \bibfield  {author} {\bibinfo {author} {\bibfnamefont {E.~H.}\ \bibnamefont
  {Lieb}}\ and\ \bibinfo {author} {\bibfnamefont {D.~W.}\ \bibnamefont
  {Robinson}},\ }\bibfield  {title} {\bibinfo {title} {The finite group
  velocity of quantum spin systems},\ }\href
  {https://doi.org/10.1007/BF01645779} {\bibfield  {journal} {\bibinfo
  {journal} {Communications in mathematical physics}\ }\textbf {\bibinfo
  {volume} {28}},\ \bibinfo {pages} {251} (\bibinfo {year} {1972})}\BibitemShut
  {NoStop}%
\bibitem [{\citenamefont {Calabrese}\ and\ \citenamefont
  {Cardy}(2006)}]{calabrese2006time}%
  \BibitemOpen
  \bibfield  {author} {\bibinfo {author} {\bibfnamefont {P.}~\bibnamefont
  {Calabrese}}\ and\ \bibinfo {author} {\bibfnamefont {J.}~\bibnamefont
  {Cardy}},\ }\bibfield  {title} {\bibinfo {title} {Time dependence of
  correlation functions following a quantum quench},\ }\href
  {https://doi.org/10.1103/PhysRevLett.96.136801} {\bibfield  {journal}
  {\bibinfo  {journal} {Phys. Rev. Lett.}\ }\textbf {\bibinfo {volume} {96}},\
  \bibinfo {pages} {136801} (\bibinfo {year} {2006})}\BibitemShut {NoStop}%
\bibitem [{\citenamefont {L{\"a}uchli}\ and\ \citenamefont
  {Kollath}(2008)}]{lauchli2008spreading}%
  \BibitemOpen
  \bibfield  {author} {\bibinfo {author} {\bibfnamefont {A.~M.}\ \bibnamefont
  {L{\"a}uchli}}\ and\ \bibinfo {author} {\bibfnamefont {C.}~\bibnamefont
  {Kollath}},\ }\bibfield  {title} {\bibinfo {title} {Spreading of correlations
  and entanglement after a quench in the one-dimensional bose--hubbard model},\
  }\href {https://doi.org/10.1088/1742-5468/2008/05/P05018} {\bibfield
  {journal} {\bibinfo  {journal} {Journal of Statistical Mechanics: Theory and
  Experiment}\ }\textbf {\bibinfo {volume} {2008}},\ \bibinfo {pages} {P05018}
  (\bibinfo {year} {2008})}\BibitemShut {NoStop}%
\bibitem [{\citenamefont {Manmana}\ \emph {et~al.}(2009)\citenamefont
  {Manmana}, \citenamefont {Wessel}, \citenamefont {Noack},\ and\ \citenamefont
  {Muramatsu}}]{manmana2009time}%
  \BibitemOpen
  \bibfield  {author} {\bibinfo {author} {\bibfnamefont {S.~R.}\ \bibnamefont
  {Manmana}}, \bibinfo {author} {\bibfnamefont {S.}~\bibnamefont {Wessel}},
  \bibinfo {author} {\bibfnamefont {R.~M.}\ \bibnamefont {Noack}},\ and\
  \bibinfo {author} {\bibfnamefont {A.}~\bibnamefont {Muramatsu}},\ }\bibfield
  {title} {\bibinfo {title} {Time evolution of correlations in strongly
  interacting fermions after a quantum quench},\ }\href
  {https://doi.org/10.1103/PhysRevB.79.155104} {\bibfield  {journal} {\bibinfo
  {journal} {Phys. Rev. B}\ }\textbf {\bibinfo {volume} {79}},\ \bibinfo
  {pages} {155104} (\bibinfo {year} {2009})}\BibitemShut {NoStop}%
\bibitem [{\citenamefont {Kim}\ and\ \citenamefont
  {Huse}(2013)}]{kim2013ballistic}%
  \BibitemOpen
  \bibfield  {author} {\bibinfo {author} {\bibfnamefont {H.}~\bibnamefont
  {Kim}}\ and\ \bibinfo {author} {\bibfnamefont {D.~A.}\ \bibnamefont {Huse}},\
  }\bibfield  {title} {\bibinfo {title} {Ballistic spreading of entanglement in
  a diffusive nonintegrable system},\ }\href
  {https://doi.org/10.1103/PhysRevLett.111.127205} {\bibfield  {journal}
  {\bibinfo  {journal} {Phys. Rev. Lett.}\ }\textbf {\bibinfo {volume} {111}},\
  \bibinfo {pages} {127205} (\bibinfo {year} {2013})}\BibitemShut {NoStop}%
\bibitem [{\citenamefont {Cheneau}\ \emph {et~al.}(2012)\citenamefont
  {Cheneau}, \citenamefont {Barmettler}, \citenamefont {Poletti}, \citenamefont
  {Endres}, \citenamefont {Schau{\ss}}, \citenamefont {Fukuhara}, \citenamefont
  {Gross}, \citenamefont {Bloch}, \citenamefont {Kollath},\ and\ \citenamefont
  {Kuhr}}]{cheneau2012light}%
  \BibitemOpen
  \bibfield  {author} {\bibinfo {author} {\bibfnamefont {M.}~\bibnamefont
  {Cheneau}}, \bibinfo {author} {\bibfnamefont {P.}~\bibnamefont {Barmettler}},
  \bibinfo {author} {\bibfnamefont {D.}~\bibnamefont {Poletti}}, \bibinfo
  {author} {\bibfnamefont {M.}~\bibnamefont {Endres}}, \bibinfo {author}
  {\bibfnamefont {P.}~\bibnamefont {Schau{\ss}}}, \bibinfo {author}
  {\bibfnamefont {T.}~\bibnamefont {Fukuhara}}, \bibinfo {author}
  {\bibfnamefont {C.}~\bibnamefont {Gross}}, \bibinfo {author} {\bibfnamefont
  {I.}~\bibnamefont {Bloch}}, \bibinfo {author} {\bibfnamefont
  {C.}~\bibnamefont {Kollath}},\ and\ \bibinfo {author} {\bibfnamefont
  {S.}~\bibnamefont {Kuhr}},\ }\bibfield  {title} {\bibinfo {title}
  {Light-cone-like spreading of correlations in a quantum many-body system},\
  }\href {https://doi.org/10.1038/nature10748} {\bibfield  {journal} {\bibinfo
  {journal} {Nature}\ }\textbf {\bibinfo {volume} {481}},\ \bibinfo {pages}
  {484} (\bibinfo {year} {2012})}\BibitemShut {NoStop}%
\bibitem [{\citenamefont {Langen}\ \emph {et~al.}(2013)\citenamefont {Langen},
  \citenamefont {Geiger}, \citenamefont {Kuhnert}, \citenamefont {Rauer},\ and\
  \citenamefont {Schmiedmayer}}]{langen2013local}%
  \BibitemOpen
  \bibfield  {author} {\bibinfo {author} {\bibfnamefont {T.}~\bibnamefont
  {Langen}}, \bibinfo {author} {\bibfnamefont {R.}~\bibnamefont {Geiger}},
  \bibinfo {author} {\bibfnamefont {M.}~\bibnamefont {Kuhnert}}, \bibinfo
  {author} {\bibfnamefont {B.}~\bibnamefont {Rauer}},\ and\ \bibinfo {author}
  {\bibfnamefont {J.}~\bibnamefont {Schmiedmayer}},\ }\bibfield  {title}
  {\bibinfo {title} {Local emergence of thermal correlations in an isolated
  quantum many-body system},\ }\href {https://doi.org/10.1038/nphys2739}
  {\bibfield  {journal} {\bibinfo  {journal} {Nature Physics}\ }\textbf
  {\bibinfo {volume} {9}},\ \bibinfo {pages} {640} (\bibinfo {year}
  {2013})}\BibitemShut {NoStop}%
\bibitem [{\citenamefont {Baier}\ \emph {et~al.}(2008)\citenamefont {Baier},
  \citenamefont {Romatschke}, \citenamefont {Son}, \citenamefont {Starinets},\
  and\ \citenamefont {Stephanov}}]{baier2008relativistic}%
  \BibitemOpen
  \bibfield  {author} {\bibinfo {author} {\bibfnamefont {R.}~\bibnamefont
  {Baier}}, \bibinfo {author} {\bibfnamefont {P.}~\bibnamefont {Romatschke}},
  \bibinfo {author} {\bibfnamefont {D.~T.}\ \bibnamefont {Son}}, \bibinfo
  {author} {\bibfnamefont {A.~O.}\ \bibnamefont {Starinets}},\ and\ \bibinfo
  {author} {\bibfnamefont {M.~A.}\ \bibnamefont {Stephanov}},\ }\bibfield
  {title} {\bibinfo {title} {Relativistic viscous hydrodynamics, conformal
  invariance, and holography},\ }\href
  {https://doi.org/10.1088/1126-6708/2008/04/100} {\bibfield  {journal}
  {\bibinfo  {journal} {Journal of High Energy Physics}\ }\textbf {\bibinfo
  {volume} {2008}},\ \bibinfo {pages} {100} (\bibinfo {year}
  {2008})}\BibitemShut {NoStop}%
\bibitem [{\citenamefont {Romatschke}(2010)}]{romatschke2010new}%
  \BibitemOpen
  \bibfield  {author} {\bibinfo {author} {\bibfnamefont {P.}~\bibnamefont
  {Romatschke}},\ }\bibfield  {title} {\bibinfo {title} {New developments in
  relativistic viscous hydrodynamics},\ }\href
  {https://doi.org/10.1142/S0218301310014869} {\bibfield  {journal} {\bibinfo
  {journal} {International Journal of Modern Physics E}\ }\textbf {\bibinfo
  {volume} {19}},\ \bibinfo {pages} {1} (\bibinfo {year} {2010})}\BibitemShut
  {NoStop}%
\bibitem [{\citenamefont {Chowdhury}\ \emph
  {et~al.}(2025{\natexlab{a}})\citenamefont {Chowdhury}, \citenamefont
  {Hartnoll}, \citenamefont {Hebbar},\ and\ \citenamefont
  {Khondaker}}]{chowdhury2025bootstrapping}%
  \BibitemOpen
  \bibfield  {author} {\bibinfo {author} {\bibfnamefont {S.~D.}\ \bibnamefont
  {Chowdhury}}, \bibinfo {author} {\bibfnamefont {S.~A.}\ \bibnamefont
  {Hartnoll}}, \bibinfo {author} {\bibfnamefont {A.}~\bibnamefont {Hebbar}},\
  and\ \bibinfo {author} {\bibfnamefont {R.}~\bibnamefont {Khondaker}},\
  }\bibfield  {title} {\bibinfo {title} {Bootstrapping transport in the
  drude-kadanoff-martin model},\ }\href@noop {} {\bibfield  {journal} {\bibinfo
   {journal} {arXiv preprint arXiv:2509.18255}\ } (\bibinfo {year}
  {2025}{\natexlab{a}})}\BibitemShut {NoStop}%
\bibitem [{\citenamefont {Chowdhury}\ \emph
  {et~al.}(2025{\natexlab{b}})\citenamefont {Chowdhury}, \citenamefont
  {Hartnoll},\ and\ \citenamefont {Hebbar}}]{chowdhury2025lieb}%
  \BibitemOpen
  \bibfield  {author} {\bibinfo {author} {\bibfnamefont {S.~D.}\ \bibnamefont
  {Chowdhury}}, \bibinfo {author} {\bibfnamefont {S.~A.}\ \bibnamefont
  {Hartnoll}},\ and\ \bibinfo {author} {\bibfnamefont {A.}~\bibnamefont
  {Hebbar}},\ }\bibfield  {title} {\bibinfo {title} {Lieb-robinson causality
  and non-fermi liquids},\ }\href@noop {} {\bibfield  {journal} {\bibinfo
  {journal} {arXiv preprint arXiv:2510.18950}\ } (\bibinfo {year}
  {2025}{\natexlab{b}})}\BibitemShut {NoStop}%
\bibitem [{\citenamefont {Israel}(1976)}]{israel1976nonstationary}%
  \BibitemOpen
  \bibfield  {author} {\bibinfo {author} {\bibfnamefont {W.}~\bibnamefont
  {Israel}},\ }\bibfield  {title} {\bibinfo {title} {Nonstationary irreversible
  thermodynamics: a causal relativistic theory},\ }\href
  {https://doi.org/10.1016/0003-4916(76)90064-6} {\bibfield  {journal}
  {\bibinfo  {journal} {Annals of Physics}\ }\textbf {\bibinfo {volume}
  {100}},\ \bibinfo {pages} {310} (\bibinfo {year} {1976})}\BibitemShut
  {NoStop}%
\bibitem [{Note1()}]{Note1}%
  \BibitemOpen
  \bibinfo {note} {The branch $t'<0$ is further suppressed by $e^{-\Omega
  t}$.}\BibitemShut {Stop}%
\bibitem [{\citenamefont {Caron-Huot}\ and\ \citenamefont
  {Moore}(2022)}]{caron2022spacelike}%
  \BibitemOpen
  \bibfield  {author} {\bibinfo {author} {\bibfnamefont {S.}~\bibnamefont
  {Caron-Huot}}\ and\ \bibinfo {author} {\bibfnamefont {G.~D.}\ \bibnamefont
  {Moore}},\ }\bibfield  {title} {\bibinfo {title} {Spacelike thermal
  correlators are almost time independent},\ }\href
  {https://doi.org/10.1103/PhysRevD.106.125015} {\bibfield  {journal} {\bibinfo
   {journal} {Phys. Rev. D}\ }\textbf {\bibinfo {volume} {106}},\ \bibinfo
  {pages} {125015} (\bibinfo {year} {2022})}\BibitemShut {NoStop}%
\bibitem [{Note2()}]{Note2}%
  \BibitemOpen
  \bibinfo {note} {The left-hand side of~\protect \textup {\hbox {\mathsurround
  \z@ \protect \normalfont (\ignorespaces \ref {eq:ac_def}\unskip \@@italiccorr
  )}} is an increasing function of $a$ that vanishes at $a=0$ (since $\Phi
  (0)=0$ and $\Phi '(0)=0$), so that $a_c$ grows monotonically with $\Omega
  /\lambda $}\BibitemShut {NoStop}%
\bibitem [{\citenamefont {Joseph}\ and\ \citenamefont
  {Preziosi}(1989)}]{joseph1989heat}%
  \BibitemOpen
  \bibfield  {author} {\bibinfo {author} {\bibfnamefont {D.~D.}\ \bibnamefont
  {Joseph}}\ and\ \bibinfo {author} {\bibfnamefont {L.}~\bibnamefont
  {Preziosi}},\ }\bibfield  {title} {\bibinfo {title} {Heat waves},\ }\href
  {https://doi.org/10.1103/RevModPhys.61.41} {\bibfield  {journal} {\bibinfo
  {journal} {Rev. of Mod. Phys.}\ }\textbf {\bibinfo {volume} {61}},\ \bibinfo
  {pages} {41} (\bibinfo {year} {1989})}\BibitemShut {NoStop}%
\bibitem [{Note3()}]{Note3}%
  \BibitemOpen
  \bibinfo {note} {Calling $G$ the Green function and $S$ the sources:
  $G^{-1}*\Psi = S_\Psi \Rightarrow \Psi =G*S_{\Psi }$, $G^{-1}*C = S_C
  \Rightarrow C=G*S_{C}$. The relation $S_\Psi = D_{\Omega }*S_C$ guarantees
  the FDT condition $\Psi =D_\Omega *C$.}\BibitemShut {Stop}%
\bibitem [{\citenamefont {Kolesnik}(2018)}]{kolesnik2018slow}%
  \BibitemOpen
  \bibfield  {author} {\bibinfo {author} {\bibfnamefont {A.~D.}\ \bibnamefont
  {Kolesnik}},\ }\bibfield  {title} {\bibinfo {title} {Slow diffusion by markov
  random flights},\ }\href {https://doi.org/10.1016/j.physa.2018.02.059}
  {\bibfield  {journal} {\bibinfo  {journal} {Physica A: Statistical Mechanics
  and its Applications}\ }\textbf {\bibinfo {volume} {499}},\ \bibinfo {pages}
  {186} (\bibinfo {year} {2018})}\BibitemShut {NoStop}%
\bibitem [{\citenamefont {Choi}\ \emph {et~al.}(2021)\citenamefont {Choi},
  \citenamefont {Mezei},\ and\ \citenamefont {S{\'a}rosi}}]{choi2021pole}%
  \BibitemOpen
  \bibfield  {author} {\bibinfo {author} {\bibfnamefont {C.}~\bibnamefont
  {Choi}}, \bibinfo {author} {\bibfnamefont {M.}~\bibnamefont {Mezei}},\ and\
  \bibinfo {author} {\bibfnamefont {G.}~\bibnamefont {S{\'a}rosi}},\ }\bibfield
   {title} {\bibinfo {title} {Pole skipping away from maximal chaos},\ }\href
  {https://doi.org/10.1007/JHEP02(2021)207} {\bibfield  {journal} {\bibinfo
  {journal} {Journal of High Energy Physics}\ }\textbf {\bibinfo {volume}
  {2021}},\ \bibinfo {pages} {207} (\bibinfo {year} {2021})}\BibitemShut
  {NoStop}%
\bibitem [{\citenamefont {Mahan}(2000)}]{mahan2000many}%
  \BibitemOpen
  \bibfield  {author} {\bibinfo {author} {\bibfnamefont {G.~D.}\ \bibnamefont
  {Mahan}},\ }\href {https://doi.org/10.1007/978-1-4757-5714-9} {\emph
  {\bibinfo {title} {Many-Particle Physics}}},\ \bibinfo {edition} {3rd}\ ed.\
  (\bibinfo  {publisher} {Springer},\ \bibinfo {address} {New York},\ \bibinfo
  {year} {2000})\BibitemShut {NoStop}%
\bibitem [{\citenamefont {Pottier}\ and\ \citenamefont
  {Mauger}(2001)}]{pottier2001quantum}%
  \BibitemOpen
  \bibfield  {author} {\bibinfo {author} {\bibfnamefont {N.}~\bibnamefont
  {Pottier}}\ and\ \bibinfo {author} {\bibfnamefont {A.}~\bibnamefont
  {Mauger}},\ }\bibfield  {title} {\bibinfo {title} {Quantum
  fluctuation-dissipation theorem: a time-domain formulation},\ }\href
  {https://doi.org/10.1016/S0378-4371(00)00523-8} {\bibfield  {journal}
  {\bibinfo  {journal} {Physica A: Statistical Mechanics and its Applications}\
  }\textbf {\bibinfo {volume} {291}},\ \bibinfo {pages} {327} (\bibinfo {year}
  {2001})}\BibitemShut {NoStop}%
\bibitem [{\citenamefont {Cugliandolo}\ \emph {et~al.}(1997)\citenamefont
  {Cugliandolo}, \citenamefont {Kurchan},\ and\ \citenamefont
  {Peliti}}]{cugliandolo1997energy}%
  \BibitemOpen
  \bibfield  {author} {\bibinfo {author} {\bibfnamefont {L.~F.}\ \bibnamefont
  {Cugliandolo}}, \bibinfo {author} {\bibfnamefont {J.}~\bibnamefont
  {Kurchan}},\ and\ \bibinfo {author} {\bibfnamefont {L.}~\bibnamefont
  {Peliti}},\ }\bibfield  {title} {\bibinfo {title} {Energy flow, partial
  equilibration, and effective temperatures in systems with slow dynamics},\
  }\href {https://doi.org/10.1103/PhysRevE.55.3898} {\bibfield  {journal}
  {\bibinfo  {journal} {Phys. Rev. E}\ }\textbf {\bibinfo {volume} {55}},\
  \bibinfo {pages} {3898} (\bibinfo {year} {1997})}\BibitemShut {NoStop}%
\bibitem [{zen()}]{zenodo}%
  \BibitemOpen
  \href@noop {} {}\bibinfo {note} {The code will be made available at
  publication.}\BibitemShut {Stop}%
\end{thebibliography}%
\appendix
\section{The quantum FDT in the time domain}\label{sec_FDT_review}
In this appendix we collect some results obtained using or deriving different FDT relations in the time domain. We split the discussion in two parts. The first presents the direct statement of the FDT in terms of a differential operator acting on $F$, which allows us to obtain the same Eq.~\eqref{eq:ac_def} for the boundary between classical diffusive hydrodynamics and the quantum regime. The second derives new results about FDT relations linking $C$ and $\Psi$.

\subsection{From $F$ to $C$ and $R$: the differential operator form}
We define
\begin{align}
    \label{eq:Rsa}
    S_{AB}(t) & =     \frac 1Z\text{Tr} \left [ {e^{-{\beta}  H} }\, A(t)      \,  B\,      \right]
    =     C_{AB}(t) +  \hbar R_{AB}''(t)\ ,
    \end{align}
with $C_{AB}(t)$ and $R_{AB}''(t)$ related to standard fluctuations and response as
\begin{subequations}
\label{eq:CRdef2}
\begin{align}
    \label{eq:Csa}
    C_{AB}(t) & = \frac 12
    \frac 1Z \text{Tr} \left [ {e^{-\beta  H} }\,\{  A(t) , \,  B\,  \}    \right ]  \ ,
 \\
    R_{AB}(t)  & = {2}i \theta(t) R_{AB}''(t) =
    \frac i \hbar \theta(t)     \frac 1Z\text{Tr} \left [ {e^{-\beta  H} }\, [ A(t) ,\,  B\,  ]    \right ] \ .
    \end{align}
\end{subequations}
The quantum FDT in frequency reads
\begin{align}\label{eq:FDT_freq}
C_{AB}(\omega) & = \cosh\!\left(\frac{\beta\hbar\omega}{2}\right)F_{AB}(\omega)\\
\hbar R''_{AB}(\omega) & = \sinh\!\left(\frac{\beta\hbar\omega}{2}\right)F_{AB}(\omega)\ .
\end{align}
In the time domain the hyperbolic functions become shift operators in imaginary time,
\begin{subequations}\label{eq:FDTop}
\begin{align}
\label{eq:FDTop_C}
C_{AB}(t)&
=\cos\!\left(\frac{\beta\hbar}{2}\frac{d}{dt}\right)F_{AB}(t),\\
\label{eq:FDTop_R}
R_{AB}(t)&
=-\frac{2}{\hbar}\theta(t)\sin\!\left(\frac{\beta\hbar}{2}\frac{d}{dt}\right)F_{AB}(t)\ .
\end{align}
\end{subequations}
In the classical limit one recovers $C_{AB}(t)=F_{AB}(t)$ and $R_{AB}(t)=-\beta\theta(t)\dot F_{AB}(t)$. Assuming
\beq\label{eq:LDP_ansatz_F}
F(x,t)\;\simeq\; e^{-\lambda\, t\,\Phi\bigl(x/(vt)\bigr)}\ ,
\eeq
Eq.~\eqref{eq:FDTop_C} gives, on the ray $x=avt$ at large times,
\beq
C(x = a v t, t) \simeq  \cos \left( \frac{\hbar\beta \lambda}{2}
\left( a  \Phi'(a) -  \Phi(a) \right)
\right) F(x,t) \ .
\eeq
The condition~\eqref{eq:ac_def} that fixes the boundary between classical and Planckian regimes is the point at which $C$ changes sign -- which automatically violates $X_{FC}(t)\simeq 1$.

\subsection{Telescoping and dyadic sum of blurred kernels}\label{Sec_dyadic}
In Ref.~\cite{pottier2001quantum}, Mauger–Pottier derived
\begin{align}
i\hbar\,R_{BA}''(t)
& =
\frac{1}{\pi}\,
C_{BA} * \operatorname{pv}\!\left(\frac{\Omega}{\sinh(\Omega t)}\right),
\\ 
C_{BA}(t)
& =
-\frac{i\hbar}{\pi}\,
R_{BA}'' * \operatorname{pv}\!\bigl(\Omega\,\coth(\Omega t)\bigr),\label{pottier}
\end{align}
where $*$ denotes convolution. We wish to relate $T\Psi_{AB}$ and $C_{AB}$, which classically coincide at equilibrium. The FDT relation in frequency reads
\begin{align}\label{eq:FDT}
\widehat\Psi_{AB}(\omega)
& =
\frac{2}{\hbar\omega}\,
\tanh\!\left(\frac{\beta\hbar\omega}{2}\right)
\widehat C_{AB}(\omega),
\\
\widehat C_{AB}(\omega)
& =
\frac{\hbar\omega}{2}\,
\coth\!\left(\frac{\beta\hbar\omega}{2}\right)
\widehat\Psi_{AB}(\omega).
\end{align}
In the time domain these become convolutions with kernels $D_T$ and $E_T$,
\begin{align}
T\,\Psi_{AB}(t)
&=
\int_{-\infty}^{\infty}dt'\;D_T(t-t')\,C_{AB}(t')
\;\equiv\; D_T * C_{AB},
\label{eq:CtoPsi}\\[4pt]
\beta\,C_{AB}(t)
&=
\int_{-\infty}^{\infty}dt'\;E_T(t-t')\,\Psi_{AB}(t')
\;\equiv\; E_T * \Psi_{AB},
\label{eq:PsitoC}
\end{align}
with
\begin{align}\label{eq:DT_ET}
D_T(t)
& =
-\frac{2T}{\pi\hbar}\,\ln\tanh\!\left(\frac{\pi T|t|}{2\hbar}\right),
\\
E_T(t)
& =
-\frac{\pi T}{2\hbar}\,
\operatorname{fp}\!\left(\frac{1}{\sinh^{2}\!\left(\frac{\pi Tt}{\hbar}\right)}\right).
\end{align}
The kernel $D_T$ is positive-definite, integrable, and decays as a Planckian exponential. $E_T$ has a non-integrable divergence at the origin, handled by the Hadamard finite part. We now derive dyadic telescoping expansions that replace both kernels by series of smooth bell-shaped functions.

\subsubsection*{Upward telescope for $D_T$}
Define
\begin{equation}\label{eq:BT}
B_T(t)
\equiv
\frac{2T}{\pi\hbar}\,
\ln\!\left(\frac{2}{1+\tanh^2\!\!\left(\dfrac{\pi T|t|}{2\hbar}\right)}\right),
\end{equation}
a smooth, positive-definite bell-shaped function with Planckian decay. One verifies the two-scale identity $D_T(t) = \tfrac{1}{2}D_{2T}(t)+B_T(t)$. Iterating $m$ times and taking $m\to\infty$ (the remainder $2^{-m}D_{2^mT}\to 0$ pointwise),
\begin{equation}\label{eq:DT_dyadic}
D_T(t) = \sum_{k=0}^{\infty}\frac{1}{2^k}\,B_{2^kT}(t),
\end{equation}
so that
\begin{equation}\label{eq:Psi_dyadic}
T\,\Psi_{AB}(t)
=
\sum_{k=0}^{\infty}\frac{1}{2^k}
\int_{-\infty}^{\infty}dt'\,B_{2^kT}(t-t')\,C_{AB}(t').
\end{equation}

\subsubsection*{Classical-subtracted telescope for $E_T$}
The non-integrable divergence of $E_T$ is regularised by subtracting the classical limit. Setting $\tilde{E}_T\equiv E_T-\delta$, Eq.~\eqref{eq:PsitoC} becomes
\begin{equation}\label{eq:PsitoC1}
\beta\,C_{AB}(t)-\Psi_{AB}(t)
=
\int_{-\infty}^{\infty}dt'\,\tilde{E}_T(t-t')\,\Psi_{AB}(t').
\end{equation}
One verifies $\tilde{E}_T(t)=2\tilde{E}_{2T}(t)+[\delta(t)-G_T(t)]$, where
\begin{align}\label{eq:FT}
G_T(t)
& =
\frac{\pi T}{2\hbar}\,
\frac{1}{\cosh^2\!\!\left(\dfrac{\pi Tt}{\hbar}\right)},
\\
\widehat{G}_T(\omega)
& =
\frac{\hbar\omega}{2T}\cdot\frac{1}{\sinh\!\left(\dfrac{\hbar\omega}{2T}\right)}
=
1-\frac{\hbar^2\omega^2}{24T^2}+\mathcal{O}(\omega^4).
\end{align}
Iterating and taking $m\to\infty$ (the remainder $2^m\tilde{E}_{2^mT}\to 0$ by the $\delta$-subtraction),
\begin{equation}\label{eq:ET_dyadic}
E_T(t)-\delta(t)
=
\sum_{k=0}^{\infty}2^k\,\bigl[\delta(t)-G_{2^kT}(t)\bigr].
\end{equation}
Substituting into~\eqref{eq:PsitoC1} yields
\begin{align}\label{eq:C_dyadic}
\beta\,C_{AB}(t)& -\Psi_{AB}(t)
=
\sum_{k=0}^{\infty}2^k\!
\left[
\Psi_{AB}(t)
-
\int_{-\infty}^{\infty}dt'\,G_{2^kT}(t-t')\,\Psi_{AB}(t')
\right].
\end{align}
Since $1-\widehat{G}_{2^kT}(\omega)\sim \hbar^2\omega^2/(24\cdot 4^k T^2)$ at small $\omega$, the $k$-th term is suppressed by $2^{-k}$ relative to the previous one, so only a few terms are needed for an accurate approximation.

\section{Fluctuation-dissipation ratios on rays}\label{sec_ratios}
Here we compute the ratios $X_{FC}=F/C$ and $X=\Psi/(\beta C)$ directly. We impose the diffusive large-deviation form $L(x,t)=t^{-1/2}\,e^{-\lambda t\,\Phi(x/vt)}$ either on $C$, blurring it as in the main text, or on $F$, as in Appendix~\ref{sec_FDT_review}, and obtain the other functions from the exact quantum FDT. At fixed $x$, the local rate and its dimensionless version are
\beq
s=\partial_t\ln L=\lambda\,[a\Phi'(a)-\Phi(a)]+O(1/t)\,,\qquad z=\frac{\hbar\beta\, s}{2}=\frac{\pi s}{2\Omega}\,.
\eeq
The blurring acts on a window of fixed width $\Omega^{-1}$, over which $L$ is a pure exponential up to $O(1/\Omega t)$ corrections. Since $\int g_\Omega(u)\,e^{su}\,du=1/\cos z$ for $|z|<\pi/2$ [equivalently, Eq.~\eqref{eq:FDTop_C} gives $C=\cos z\,F$ on a pure exponential], in both cases one finds, as $t\to\infty$ at fixed $a$ and for $z<\pi/2$,
\beq\label{eq:ratios}
X_{FC}\to\frac{1}{\cos z}\,,\qquad X\to\frac{\tan z}{z}\,.
\eeq
The condition $z=\pi/2$, i.e.\ $s=\Omega$, coincides with Eq.~\eqref{eq:ac_def} for $a_c$.

In the classical region, $a<a_c$, both choices give the same ratios, independent of $t$ and equal to $1+O(z^2)$; they are not exactly one, and they grow on approaching $a_c$. The quantitative criterion for classicality is $z\ll1$. In the Gaussian diffusive regime $z=\frac{\pi}{2}(a/a_c)^2$, so that, e.g., $X_{FC}\le1+\varepsilon$ for $a/a_c\lesssim(8\varepsilon/\pi^2)^{1/4}$.

\begin{figure}[h!]
\centering
\includegraphics[width=0.65\linewidth]{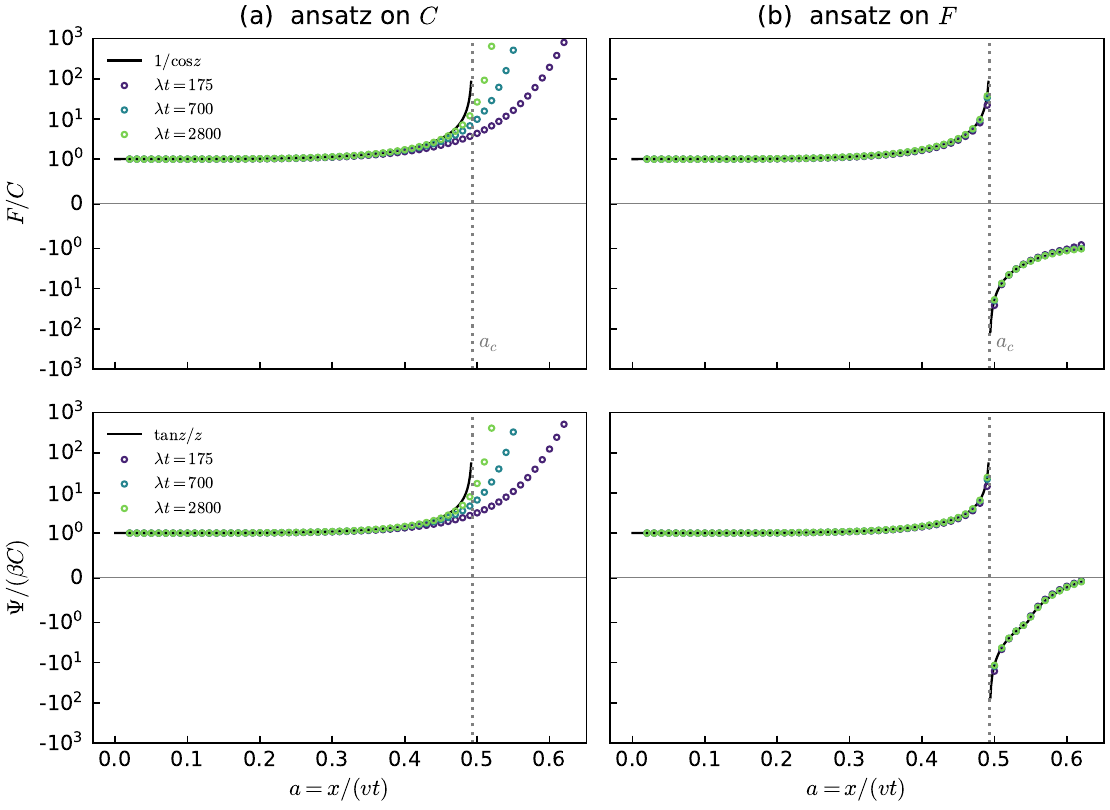}
\caption{$F/C$ (top) and $\Psi/(\beta C)$ (bottom) on rays $x=avt$ for the telegraph rate function $\Phi(a)=2(1-\sqrt{1-a^2})$, with the large-deviation form imposed on $C$ (a) or on $F$ (b). Parameters $\lambda=1.75$, $\beta=6$, giving $a_c=0.49$ (dotted line). Symbols: numerics at increasing $\lambda t$. Lines: $1/\cos z$ and $\tan z/z$ [Eq.~\eqref{eq:ratios}]. Symmetric-log vertical scale.\label{fig:ratios}}
\end{figure}

In the quantum region, $a>a_c$, the classical reading breaks down in a way that depends on which function carries the large-deviation form: starting from $C$, $F$ and $\Psi$ grow exponentially with respect to $C$; starting from $F$ (or from $\Psi$), $C$ changes sign. In no case can the correlators be read as classical up to small deviations. The boundary $a_c$ is thus the same whichever function carries the hydrodynamic form, and it becomes sharp as $t\to\infty$, as shown by the convergence of the numerical data in Fig.~\ref{fig:ratios}.

\section{Phenomenological hydrodynamic theories of heat transport}
\label{sec_hydro_theories}
In this appendix we summarise the classification of phenomenological hydrodynamic theories of heat transport reviewed by Joseph and Preziosi in Ref.~\cite{joseph1989heat}, which provides the physical motivation for the equations studied in the main text. All start from a conserved quantity $n(x,t)$ with continuity equation
\beq\label{App_Continuity}
\partial_t n(x,t) + \nabla j(x,t) = 0\,,
\eeq
and differ in the constitutive relation between current and density.

The simplest is the Fick–Fourier law, $j = -D\nabla n$, which, combined with continuity, yields the diffusion (or heat) equation $\partial_t n = D\Delta n$. It describes transport on long time and length scales but carries a well-known pathology: solutions spread with infinite speed.

Cattaneo's modification removes this by allowing the current to relax to the Fick value over a finite time $\tau$,
\beq\label{App_Cattaneo}
\tau\,\partial_t j + j = -D\,\nabla n\,,
\eeq
which can be written in integral form as $j(x,t) = -(D/\tau)\int_{-\infty}^{t}\!dt'\,e^{-(t-t')/\tau}\,\nabla n(x,t')$. Combined with continuity, this gives the telegraph equation
\beq
\partial_t^2 n + 4\lambda\,\partial_t n = v^2\,\Delta n\,,
\eeq
with $4\lambda=1/\tau$ and $v^2=D/\tau$. It interpolates between a wave equation at short times and diffusion at long times, with effective constant $D=v^2/(4\lambda)$, and possesses a sharp light cone at $|x|=vt$.

The Jeffreys generalisation supplements Cattaneo's law with a term proportional to the gradient of the time derivative of the density:
\beq
\tau\,\partial_t j + j = -D\,\nabla n - \tau\kappa\,\nabla\partial_t n\,,
\eeq
yielding
\beq
\partial_t^2 n + 4\lambda\,\partial_t n = v^2\,\Delta n + \kappa\,\partial_t\Delta n\,,
\eeq
a telegraph equation with an additional diffusive term that smooths the ballistic-to-diffusive crossover. In terms of a memory kernel $Q(s)$, the Jeffreys law corresponds to $Q(s) = \kappa\,\delta(s) + (\kappa_2/\tau)\,e^{-s/\tau}$, and the diffusion constant is $D = \kappa + \kappa_2$; comparing with the differential form above, $\kappa_2=D-\kappa$, so that the total diffusion constant is the same $D=v^2/(4\lambda)$ as in the main text.

In the main text we study the consequences of the quantum FDT on the solutions of each of these equations, treated as effective descriptions of two-point correlation functions in many-body quantum systems at long times and large distances. The hierarchy -- from diffusion to telegraph to diffusive-telegraph -- provides increasingly refined models of how finite propagation speed and memory effects shape transport, and in each case the interplay between the hydrodynamic decay rate and the Matsubara frequency $\Omega = \pi/(\hbar\beta)$ controls whether classical hydrodynamics or quantum effects prevail.

\end{document}